\documentclass[12pt,oneside]{article}
\usepackage[a4paper,margin = 2cm]{geometry}
\usepackage{tikz}
\usepackage[OT1]{fontenc}
\usepackage{times}
\usepackage{amsthm}
\usepackage{amssymb}
\usepackage{amsfonts}
\usepackage{setspace}
\usepackage{newtxtext,newtxmath}
\usepackage{graphicx}
\usepackage{rotating}
\usepackage{authblk}
\usepackage{xcolor}
\definecolor{lightblue}{rgb}{0.85,0.00,0.99}

\usepackage{color}
\usepackage[round]{natbib}
\usepackage{bm}
\usepackage{bigints}
\usepackage{multirow}
\usepackage{float}
\usepackage{longtable}
\usepackage{lscape}
\usepackage{setspace}
\usepackage{mathtools}
\usepackage{multicol}
\usepackage{caption, subcaption}
\usepackage{makecell}
\usepackage{pdflscape}
\usepackage{pdfpages}
\usepackage[colorlinks,citecolor=blue,urlcolor=blue]{hyperref}

\usetikzlibrary{shapes,decorations,arrows,calc,arrows.meta,fit,arrows,shapes.arrows,
shapes.geometric,shapes.multipart,decorations.pathmorphing,positioning}
\tikzset{
    -Latex,auto,node distance =1 cm and 1 cm,semithick,
    state/.style ={ellipse, draw, minimum width = 0.7 cm},
    point/.style = {circle, draw, inner sep=0.04cm,fill,node contents={}},
    bidirected/.style={Latex-Latex,dashed},
    el/.style = {inner sep=2pt, align=left, sloped}
}

\usepackage{bbm}
\newcommand{\hide}[1]{}

\renewcommand{\P}{\mbox{P}}

\newcommand{\E}{\mathrm{E}}

\newtheorem{proposition}{Proposition}

\title{On the magnitude, sign and ranking of recanting-twin path-specific effects}
\author[1]{Tran Trong Khoi Le}
\author[2]{Pham Hien Trang Tu}
\author[3]{Nhat Long Ngo}
\author[1,*]{Tat-Thang Vo}

\affil[1]{\small EPILOGY, Institut Mondor of Biomedical Research,
INSERM U955, University Paris Est Créteil, France}

\affil[2]{\small Division of Pharmacoepidemiology and Clinical Pharmacology,
Utrecht University, The Netherlands}

\affil[3]{\small Hasselt University, Belgium}

\definecolor{lightblue}{rgb}{0.85,0.00,0.99}

\begin{document}
\maketitle
\begin{abstract}
The framework of recanting twin path-specific effects has recently been propose to address the issue of intermediate confounding in causal mediation analysis, enabling the decomposition of the average treatment effect into identifiable fine-grained path-specific effects (PSEs). An open question, however, is the extent to which recanting-twin PSEs reflect the direction and relative magnitude of their corresponding natural PSEs. In this paper, we systematically characterize the behaviors of the recanting-twin PSEs in terms of magnitude, sign and ranking, benchmarking against natural PSEs. To achieve this, we first derive non-parametric, identifiable upper and lower bounds for the absolute difference between recanting-twin and natural PSEs in binary outcome settings. These bounds provide a practical way to assess whether discrepancies in magnitude between the two types of effects are substantial,
thereby facilitating the use of recanting-twin PSEs as informative approximations of their natural counterparts. A simulation study is then conducted to evaluate frequency of disagreements in sign and ranking between recanting-twin and natural effects in non-linear, non-monotonic scenarios. Across four scenarios and 320 numerical settings, we find that disagreement ranges from 15\% to 56\% for ranking and from 5\%
to 43\% for sign. Disagreement is strongly influenced by the non-observable correlation between counterfactual values of the intermediate confounder, suggesting that caution is needed when interpreting the magnitude and ranking of recanting-twin PSEs, and motivating future work to identify conditions under which the two types of PSEs yield consistent conclusions when non-monotonicity presents.
\end{abstract}
\def\thefootnote{*}\footnotetext{Corresponding author: Tat-Thang Vo. EPILOGY, Institut Mondor of Biomedical Research, INSERM U955, University Paris Est Créteil, France. Email: tat-thang.vo@u-pec.fr.}\def\thefootnote{\arabic{footnote}}

\section{Introduction}
Causal mediation analysis refers to a class of statistical methods used to understand how and why a treatment or exposure affects an outcome through intermediate variables known as mediators \citep{robins1992,Pearl01}. In the simplest setting of one mediator, the total effect can be decomposed into a natural direct effect and a natural indirect effect. These effects are non-parametrically identified under two crucial assumptions: (i) all confounders of the relationships between the exposure, the mediator and the outcome are measured, and (ii) none of the mediator-outcome confounders are themselves affected by the exposure, i.e., there are no intermediate confounders \citep{Avin_2005}. 

When multiple causally related mediators are involved, earlier mediators may act as intermediate confounders for later ones, making natural indirect effects through these later mediators unidentifiable from observed data. As an alternative, attention often shifts to decomposing the total effect into multiple path-specific effects (PSEs) that correspond to distinct causal pathways. For instance, with two mediators, the treatment effect can be attributed to four causal paths involving one, both or neither of the mediators. Several methods for path analysis have been proposed, typically by extending the standard natural direct and indirect effects into natural path-specific effects, using nested counterfactuals \citep{vanderweele2014effect, steen2017medflex, vo2022longitudinal}. A central challenge, however, is that defining certain natural PSEs may involve a recanting witness -- that is, a counterfactual variable that plays conflicting roles across different causal paths \citep{Avin_2005}. These natural PSEs are generally not identifiable from the observed data, even under a non-parametric structural equation model with independent errors that allows for strong cross-world independencies \citep{Pearl_2001}.  

To address this issue, \cite{diaz2022causal} recently introduced a new class of PSEs that are defined based on random draws from the distribution of the recanting witness. These so-called recanting-twin PSEs are instances of randomized interventional effects that were originally proposed by \citet{didelez2012direct}.  Unlike natural PSEs, recanting-twin PSEs are identifiable from the observed data, making them viable alternatives for measuring the exposure effect mediated through specific causal pathways. Importantly, recanting-twin PSEs satisfy the path-specific sharp null criteria, in the sense that they take null values whenever no exposure effect is mediated via the corresponding paths. While recanting-twin and natural PSEs may differ in magnitude when the paths are non-null, they align in direction under monotonicity, i.e., when the mediated effect along each path is in the same direction for all individuals in the population \citep{diaz2022causal, vo2024recanting}. 

In complex settings where monotonicity is violated, the relative behavior of recanting-twin and natural PSEs remains underexplored. In particular, it is unclear how frequently these two estimands disagree in direction at the population level when individual-level path-specific effects vary in sign across individuals. Another open question concerns the reliability of using recanting-twin estimands to rank the relative importance of different causal pathways. More specifically, when the recanting-twin effect along one pathway is larger in magnitude than that along another, can this be interpreted as evidence that the corresponding natural effect is also larger? This question is critical in applications where the ranking of causal pathways inform targeted interventions or guide resource allocation, for example, prioritizing a health policy that acts through the most impactful pathway. An incorrect ranking could lead to misguided resource allocation or misplaced policy emphasis.

In this paper, we evaluate the behavior of recanting-twin PSEs, specifically in terms of their magnitude, sign, and ranking in non-monotonic settings, using natural PSEs as a benchmark. Section 2 begins with a brief review of both types of PSEs in the context of two mediators. In Section 3.1, we show analytically that, under linear outcome models with no mediator-intermediate confounder interaction, natural and recanting-twin PSEs coincide exactly. This equivalence also extends to log-linear outcome models, provided all effects are defined on the multiplicative scale. To assess inconsistency beyond these parametric settings, Section 3.2 establishes nonparametric and identifiable upper and lower bounds for the absolute difference between the two estimands. These bounds provide a practical way to assess when recanting-twin PSEs can offer informative approximations of the magnitude of their natural counterparts. Section 3.3 presents a simulation study to empirically assess the concordance between natural and recanting-twin PSEs in terms of sign and ranking under settings that incorporate both nonlinearity and non-monotonicity. Finally, Section 4 concludes with a discussion of the implications of these findings for applied mediation analysis.

\section{Effect decomposition under intermediate confounding}
Consider a mediation analysis that aims to explain the effect of a binary exposure $A$ on an outcome $Y$ via two sets of intermediate variables $Z$ and $M$. The relationship between these variables are visualized by the causal diagram depicted in Figure \ref{fig:1}. Throughout this paper, we assume that individual-level data on $(W,A,Z,M,Y)$ are generated according to a non-parametric structural equation model with independent errors (NPSEM-IE) \citep{Pearl_2001}.  Let $Y(a)$ be the counterfactual value of $Y$ under a hypothetical intervention that set $A=a$. The total causal effect of the exposure $A$ on the outcome $Y$ can then be defined as
$
\psi = \E\{ Y(1) - Y(0) \}.
$
Our objective is to decompose $\theta$ into four distinct components, each corresponding to a specific causal pathway through which the effect of $A$ propagates to $Y$, i.e.:
\begin{itemize}
    \item $P_1: A \rightarrow Y$, the direct effect;
    \item $P_2: A \rightarrow Z \rightarrow Y$, the indirect effect through $Z$ only;
    \item  $P_3: A \rightarrow Z \rightarrow M \rightarrow Y$, the indirect effect through both $Z$ and $M$; and
    \item  $P_4: A \rightarrow M \rightarrow Y$, the indirect effect through $M$ only.
\end{itemize}
\subsection{Natural path-specific effects}
Let $Z(a), M(a,z)$ and $Y(a,z,m)$ denote the counterfactual values of $Z, M$ and $Y$ that could be observed under hypothetical interventions that set $A=a$; $(A,Z)=(a,z)$ and $(A,Z,M)=(a,z,m)$, respectively. The effect operating through path $P_ j;j=1,\ldots,4$ can then be defined as \[\psi_{P_j}^{ne}=\E(Y_{S_{j-1}}) - \E(Y_{S_j}),\] 
where $Y_{S_j}$'s are the nested counterfactuals:
\begin{equation}
\label{Y_Sj}
    \begin{aligned}
    &Y_{S_0}  = Y(a, Z(a), M(a, Z(a))),  \qquad&
    &Y_{S_1}  = Y(a^{*}, Z(a), M(a, Z(a))), \\
    &Y_{S_2}  = Y(a^{*}, Z(a^{*}), M(a, Z(a))),  \qquad&
    &Y_{S_3}  = Y(a^{*}, Z(a^{*}), M(a, Z(a^{*}))), \\
    &Y_{S_4}  = Y(a^{*}, Z(a^{*}), M(a^{*}, Z(a^{*}))), 
    \end{aligned}
\end{equation}
for $a,a^*=0,1$. These effects are referred to as \textit{natural path-specific effects} (PSEs), since the levels of the mediator and intermediate confounder in these effects are fixed at values naturally observed under different interventions on the exposure. Under standard assumptions encoded in the NPSEM-IE \ref{fig:1}, the distributions of $Y_{S_0}, Y_{S1}, Y_{S_3}$ and $Y_{S_4}$ can be non-parametrically identified from the observed data \citep{VanderWeele_2014}. In contrast, identification of $Y_{S_2}$ requires $Z$ to simultaneously behave as though $A=a$ in one part and $A=a^*\ne a$ in the other part. This contradiction renders $Y_{S_2}$, and thus the path-specific effects $P_2$ and $P_3$, generally non-identifiable and non-estimable without invoking stronger assumptions about the data-generating process. In the current literature, $Z$ is often referred to as a \textit{recanting witness} \citep{Avin_2005}.

\subsection{Recanting twin path-specific effects}
To address the above concern, \citet{diaz2022causal} and \citet{vo2024recanting} recently introduced a new concept called \textit{recanting twin} and based on it, constructed a novel class of path-specific effects that are identifiable from the observed data. Specifically, consider a random draw $T(a)$ from the distribution of $Z(a)$  conditional on $W$. This random draw, referred to as the \textit{recanting twin} of $Z(1-a)$, replaces the recanting witness $Z(a)$ in the original $Y_{S_j}$ to create a new set of four interventional potential outcomes, i.e.:
\begin{align*}
  Y_{S_1}' &= Y(a^*, Z(a), M(a, T(a))),\qquad&
  Y_{S_2}'' &= Y(a^*, T(a^*), M(a, Z(a))),\\
  Y_{S_2}' &= Y(a^*, Z(a^*), M(a, T(a))),\qquad&
  Y_{S_3}'' &= Y(a^*, T(a^*), M(a, Z(a^*))),
\end{align*}
Based on this, the average treatment effect can be decomposed into five components, i.e. $\psi = \psi^{rt}_{P_1} + \psi^{rt}_{P_2} + \psi^{rt}_{P_3} + \psi^{rt}_{P_4} +
  \psi_{P_2\vee P_3},$ where:
\begin{align*}
  &\psi_{P_1}^{rt} = \E(Y_{S_0} - Y_{S_1});  &  
  \psi_{P_2}^{rt} = \E(Y_{S_1}' - Y_{S_2}');\\
  &\psi_{P_3}^{rt} = \E(Y_{S_2}'' - Y_{S_3}'');& 
  \psi_{P_4}^{rt} = \E(Y_{S_3}  - Y_{S_4});
\end{align*}
are the path-specific effects, and:
\[ \psi_{P_2\vee P_3} = \E(Y_{S_1} - Y_{S_1}' + Y_{S_2}' - Y_{S_2}'' +
  Y_{S_3}'' - Y_{S_3})\] 
is a parameter that identifies the presence of intermediate confounding, in the sense that $\psi_{P_2\vee P_3}=0$ when one of the three arrows $A\rightarrow Z$, $Z\rightarrow M$ and $Z\rightarrow Y$ is absent on the causal diagram, so that $Z$ is not an intermediate confounders. Details on the identifiability and non-parametric estimation strategies of $\psi_{P_j}^{rt}$ and $\psi_{P_2\vee P_3}$ can be found elsewhere \citep{vo2024recanting}.

The recanting-twin PSEs possess several important properties. First, these effects are null whenever the corresponding paths play no role in mediating the exposure effect on the outcome for any individual in the population. As such, they can be used to assess the presence or absence of all four causal pathways \citep{diaz2022causal}. Second, recanting-twin PSEs are equal to the natural PSEs when $Z$ is not an intermediate confounder. In practice, if the null hypothesis of $\psi_{P_2\vee P_3}=0$ is rejected and there is compelling substantive knowledge suggesting that $Z$ is unlikely to be an intermediate confounder, it is reasonable to interpret recanting-twin PSEs as natural PSEs \citep{vo2024recanting}. In contrast, when $Z$ is potentially an intermediate confounder, the two types of effect are no longer equivalent. However, they may still share the same sign if the mediated effect through each pathway points in the same direction for all individuals in the population, so-called monotonicity \citep{diaz2022causal}. As an example, if the individual-level effect mediated via path $P_2: A\rightarrow Z \rightarrow Y$ has the same direction for all subjects in the population, i.e.:
\[
Y(a_1,Z(a_2), M(a_4,Z(a_3))) < Y(a_1,Z(a_2'), M(a_4,Z(a_3))) \quad \text{whenever} \quad a_2<a_2'
\]
then $\psi_{P_2}=\E(Y_{S_1}'-Y_{S_2}') <0$ for $a < a^*$. This monotonicity condition ensures that recanting-twin effects remain informative about the direction (positive or negative) of pathway-specific impacts. 

However, monotonicity may not hold in practice, i.e. the impact of the exposure on the outcome via a specific path may be in opposite direction for different individuals. In such cases, one may still hope that natural and recanting-twin effects agree in sign at the population level; for example, $\psi_{P_2}^{rt}<0$ when $\psi_{P_2}^{ne}<0$. A related question concerns whether the ranking of natural PSEs is preserved among recanting-twin PSEs. For instance, if $0<\psi_{P_2}^{ne}<\psi_{P_3}^{ne}$, so that path $P_3$ plays a larger role than path $P_2$ in explaining the treatment effect, it is unclear whether this ordering is maintained under the recanting-twin framework, that is, whether $0<\psi_{P_2}^{rt}< \psi_{P_3}^{rt}$. We examine these sign and ranking properties of recanting-twin PSEs in the following section.     

\section{Relative behaviors of natural and recanting-twin effects}
\subsection{Coincidence of natural and recanting-twin effects in standard linear settings}
As a starting point, we analytically examine the behavior of recanting-twin PSEs in the simplest setting, where the outcome follows a linear model. This analysis is particularly relevant because linear mediation models remain widely used in applied research, especially in psychology. As shown below, assuming linearity allows for closed-form expressions of both natural and recanting-twin effects. These expressions help illuminate the factors that influence the sign and relative magnitude of recanting-twin PSEs in comparison to natural PSEs. Understanding these factors in the linear case provides insight into the potential sources of inconsistency between the two types of PSEs in more complex, nonlinear settings.

Assume the following outcome model:
\[
\E(Y(a,z,m)\mid W) = \gamma_0^{a,W} + \gamma_1^{a,W}z + \gamma_2^{a,W}m +  \gamma_3^{a,W}zm 
\]
Under this model, the natural effect and recanting-twin effect that describe the path $P_2$ and $P_3$ can be written in closed-form expressions as (Appendix A):
\begin{align*}
    \psi_{P_2}^{ne} &=\gamma_1^{a,W}~ \mathsf{ATE}_Z^W + \gamma_3^{a,W}~\E[Z(a^*)M(a^*)-Z(a)M(a^*)\mid W]\\[5pt]
    \psi_{P_2}^{rt}
    &=\gamma_1^{a,W}~\mathsf{ATE}_Z^W + \gamma_3^{a,W}~\mathsf{ATE}_Z^W~\E[M(a^*)\mid W]  \\
    \psi_{P_3}^{ne} &=\gamma_2^{a,W}~\mathsf{NIE}_M^W
    + \gamma_3^{a,W}~\bigg\{\E[Z(a)M(a^*)\mid W] 
    - \sum_zz~\E[M(a^*,z)\mid W] ~\P[Z(a)=z\mid W]\bigg\}\\
     \psi_{P_3}^{rt} &=
     \gamma_2^{a,W}~\mathsf{NIE}_M^W + \gamma_3^{a,W}~\mathsf{NIE}_M^W ~\E[Z(a)\mid W]
\end{align*}
where $\mathsf{ATE}_Z^W = \E[Z(a^*)-Z(a)\mid W]$ is the average treatment effect of $A$ on $Z$ given $W$, and $\mathsf{NIE}_M^W=\E\{M(a^*) - M(a^*,Z(a))\mid W\}$ is the natural indirect effect of $A$ on $M$ via $Z$ given $W$. These expressions suggest that when there is no interaction between $Z$ and $M$ in the outcome model $(\gamma_3^{a,W}=0)$, natural and recanting-twin PSEs coincide exactly and both are identifiable from the observed data under appropriate independence assumptions imposed by the associated NPSEM-IE. In contrast, when $\gamma_3^{a,W}\ne0$, the two types of effect may diverge. The extent of this discrepancy depends on various parameters in the data-generating process, including among others, 
\begin{itemize}
    \item [(i)] the strength of the interaction effect $\gamma_3^{a,W}$,
    \item [(ii)] the strength of the effect of $A$ on $Z$, i.e. $\mathsf{ATE}_Z^W$, 
    \item [(iii)] the direct effect of $Z$ on $M$ (which determines the magnitude of $\E\{M(1,z)\mid W\}$ and $\mathsf{NIE}_M^W$)
    \item [(iv)] the correlation between $Z(0)$ and $Z(1)$ (which determines the magnitude of the cross-world term $\E[Z(a)M(a^*)\mid W]$)
\end{itemize}
These factors interact in complex ways, and their interplay can lead to disagreement between the natural and recanting-twin PSEs, both in terms of magnitude, sign or relative ranking. In Appendix A, we further prove that these remarks also hold under a log-linear outcome model, provided that the PSEs are defined on the multiplicative scale.

It is worth noting that the absence of $Z-M$ interaction in the linear outcome model is a sufficient (but not necessary) condition for the coincidence of recanting-twin and natural PSEs. Under additional (linear) constraints on the data generating process, it is possible for the two effects to align even when $Z$ modifies the effect of $M$ on $Y$. As an example, consider the NPSEM-IE that assumes additionally standard linear models for $Z$ and $M$, i.e.:
\begin{align*}
\label{eq:FLM}
    \begin{aligned}
    Z(a) &= \alpha_0^W + \alpha_{1}^Wa + \epsilon_Z,   \\
    M(a,z) &= \beta_0^{a,W} + \beta_{1}^{a,W}z + \epsilon_M,   \\
    Y(a,z,m) &= \gamma_0^{a,W} + \gamma_{1}^{a,W}z + \gamma_{2}^{a,W}m +\gamma_{3}^{a,W}zm + \epsilon_Y
    \end{aligned}
\end{align*}
where the mean-zero random errors $\epsilon_Z$, $\epsilon_M$ and $\epsilon_Y$ are pairwise independent. Such a linear structural equation model imposes the strong assumption that the individual-level effect of $A$ on $Z$ is constant in the subgroup of patients having the same value of $W$, i.e. $Z(1)-Z(0)=\alpha_1^W$, which allows the identifiability and equality of natural PSEs and recanting-twin PSEs. 
A formal proof is provided in Appendix A.


\subsection{Non-parametric bounds for the absolute difference between natural and recanting-twin effects in general settings}
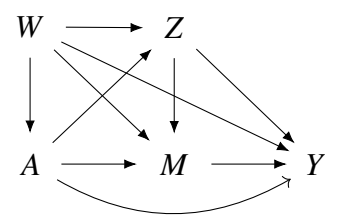
\begin{figure}[h]
    \centering
        \begin{tikzpicture}[node distance =1 cm and 1 cm]
        \node[state,draw = none] (a) at (0,0) {$A$};
        \node[state,draw = none] (m) [right =of a] {$M$};
        \node[state,draw = none] (y) [right =of m] {$Y$};
        \node[state,draw = none] (z) [above =of m] {$Z$};
        \node[state,draw = none] (w) [above =of a] {$W$};
        \path (a) edge (m);\path (m) edge (y);
        \draw [->] (a) to [out=-30, in=-150] (y);
        \path (a) edge (z);\path (z) edge (y);
        \path (z) edge (m);
        \path (w) edge (a);\path (w) edge (z);\path (w) edge (m);\path (w) edge (y); 
        \end{tikzpicture}   
    \caption{Causal diagram. $W$ denotes baseline covariates, $A$ the binary exposure, $Z$ the intermediate confounders, $M$ the mediators, and $Y$ the outcome.}
    \label{fig:1}
\end{figure}

\paragraph{A lower bound of the total difference} Beyond the full linear structural equation model, the difference between the two types of PSEs can not generally be expressed in closed-form. However, a lower bound on such difference can be derived based on the fact that the effect mediated through both paths $P_2$ and $P_3$ are identifiable by both PSE systems. More specifically, 
\[
\psi_Z^{ne}:=\psi_{P_2}^{ne} + \psi_{P_3}^{ne} = \E(Y_{S_1})-\E(Y_{S_3})
\]
reflect the natural indirect effect via $Z$, which does not involve the ill-identified $Y_{S_2}$ and thus is estimable from data under the NPSEM-IE encoded in the causal graph \ref{fig:1}. Similarly,
\[
\psi_Z^{rt}:=\psi_{P_2}^{rt} + \psi_{P_3}^{rt} = \E(Y_{S_1}')-\E(Y_{S_2}') + \E(Y_{S_2}'') - \E(Y_{S_3}'')=\E(Y_{S_1}') - \E(Y_{S_3}'')
\]
where the second equality holds since $\E(Y_{S_2}') = \E(Y_{S_2}'')$ under the same NPSEM-IE (see \citet{vo2024recanting} for a formal proof). 

\begin{figure}[H] 
        \centering
        \begin{subfigure} {0.8\textwidth}
        \centering
            \includegraphics[width=0.85\linewidth]{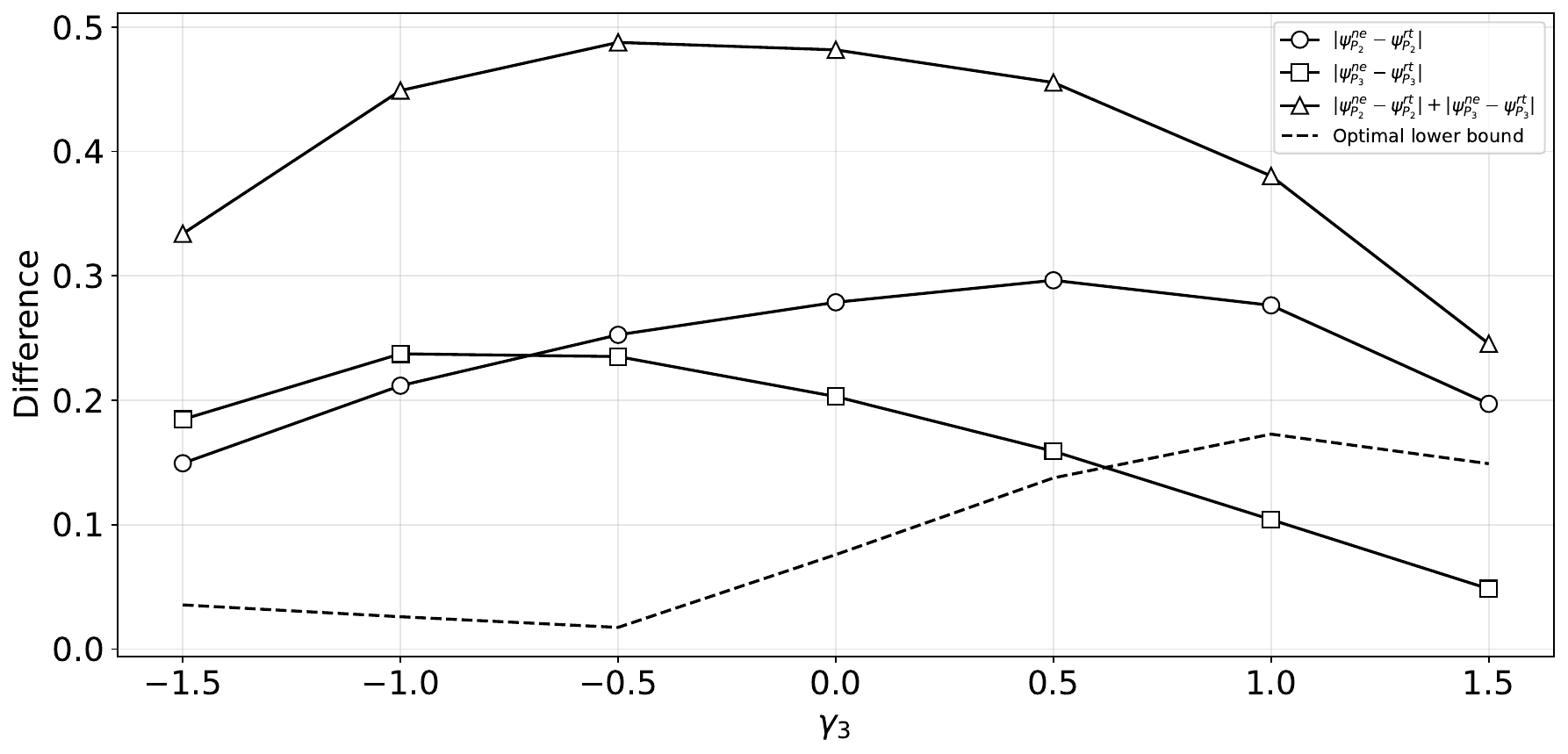}
            \caption{}            
            \label{fig:2a}
        \end{subfigure}
        
        \begin{subfigure} {0.8\textwidth}
        \centering
            \includegraphics[width=0.85\linewidth]{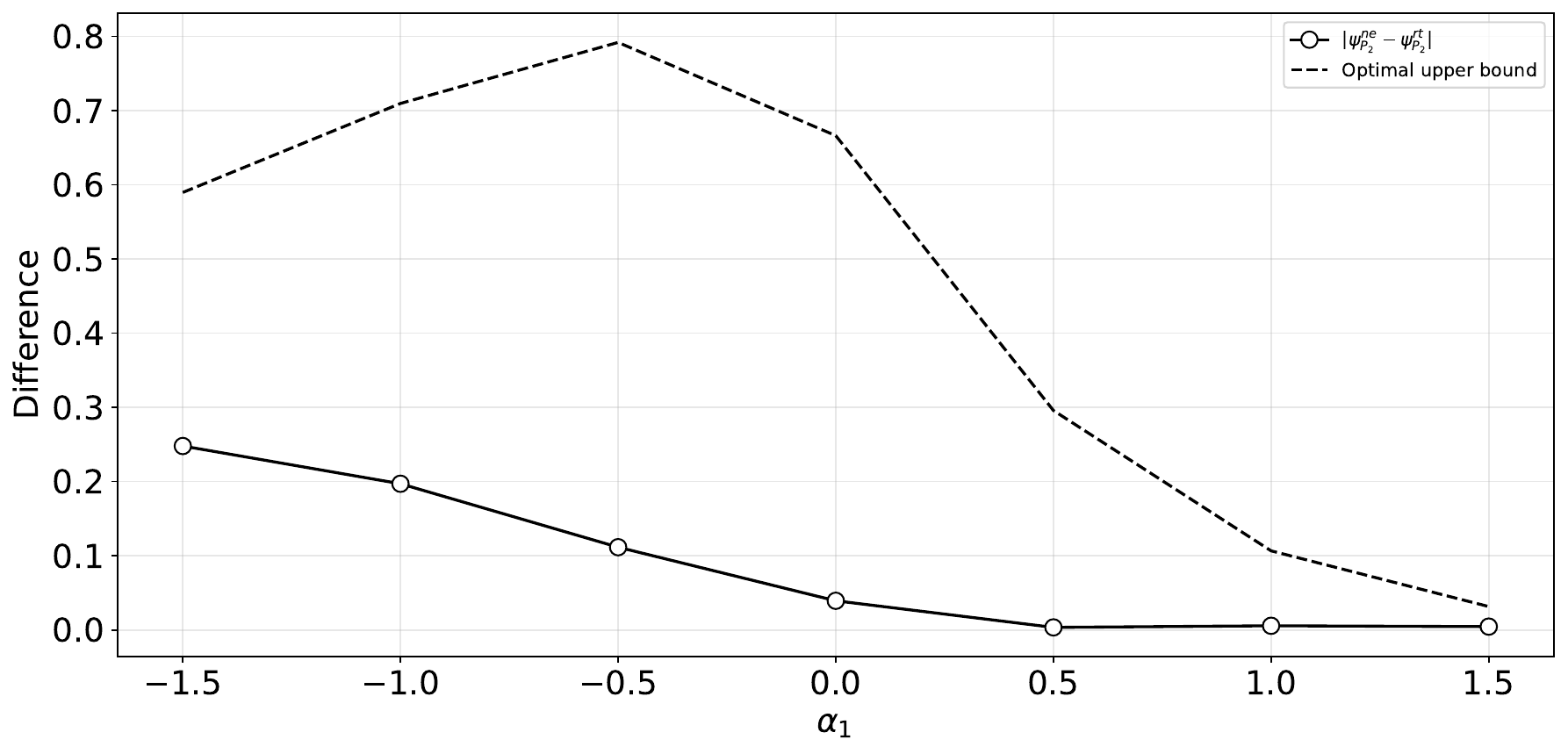}
            \caption{}
            \label{fig:2b}
        \end{subfigure}
        
        \begin{subfigure} {0.8\textwidth}
        \centering
            \includegraphics[width=0.85\linewidth]{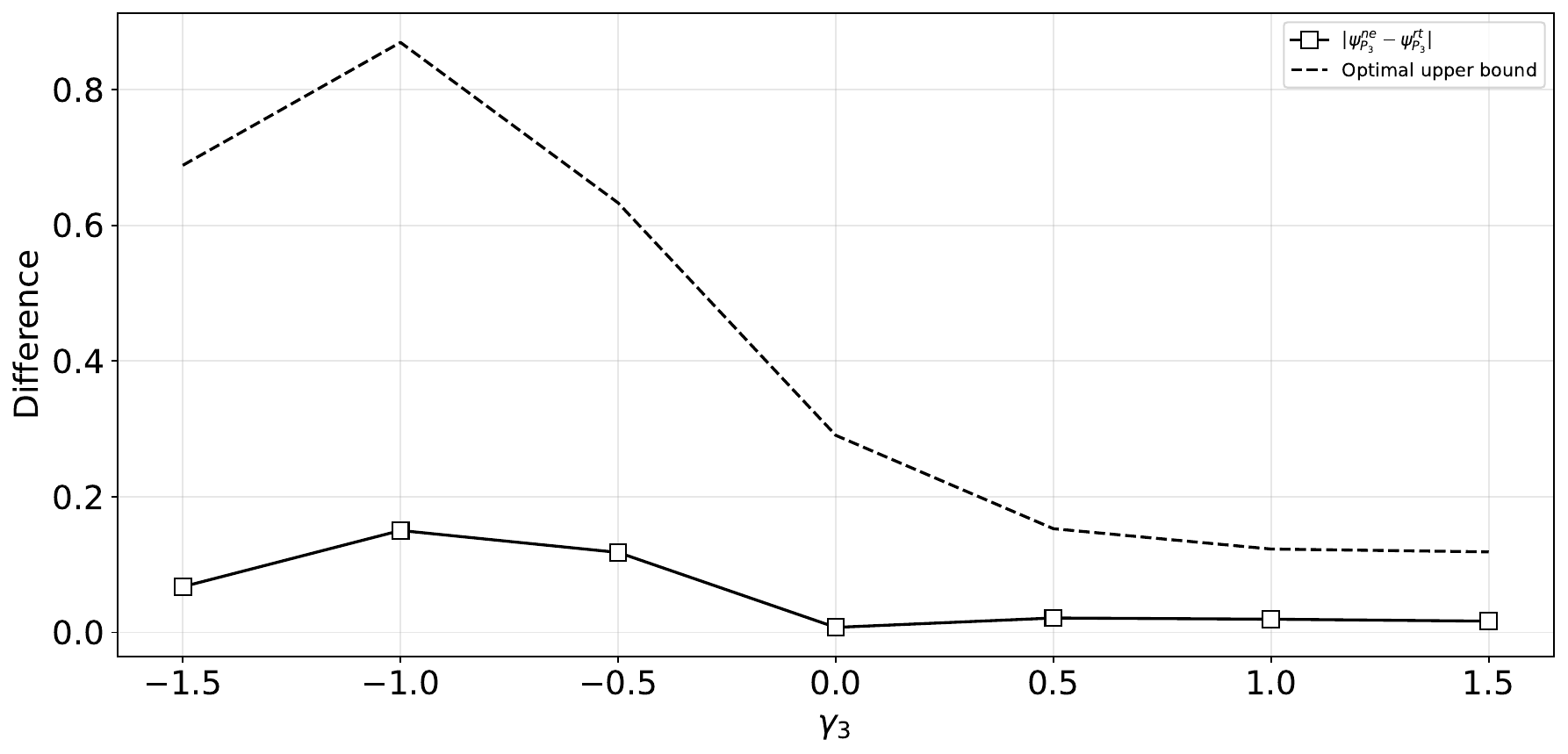}
            \caption{}
            \label{fig:2c}
        \end{subfigure}
    \caption{A toy example illustrating the informativeness of the proposed bounds using numerically simulated data. Details of the simulation setups are provided in Appendix B. Each bound is illustrated using a different simulation setup. Specifically, we explore the sensitivity of each bound to various parameters of the data-generating mechanism, then select the parameter that produces the most pronounced variation in the bound. The figures visualize the changes of the oracle values of the bound and corresponding absolute difference over different values of the selected parameter. (a) The lower bound for $|\psi_{P_2}^{ne} - \psi_{P_2}^{rt}| + |\psi_{P_3}^{ne} - \psi_{P_3}^{rt}|$ is informative for positive values of $\gamma_3$, which determines the magnitude of the interaction between $M$ and $Z$ on $Y$. (b) The upper bound for $|\psi_{P_2}^{ne} - \psi_{P_2}^{rt}|$ is informative for positive values of $\alpha_1$, which determines the magnitude of the effect of $A$ on $Z$. (c) The upper bound for $|\psi_{P_3}^{ne} - \psi_{P_3}^{rt}|$ is informative for positive values of $\gamma_3$, which determines the magnitude of the interaction between $M$ and $Z$ on $Y$.}
    \label{fig:2}
    \end{figure}

Note that if $\psi_Z^{ne} \ne \psi_Z^{rt}$ then it is necessary true that either $\psi_{P_2}^{ne} \ne \psi_{P_2}^{rt}$ or $\psi_{P_3}^{ne} \ne \psi_{P_3}^{rt}$, or both.  Thus, if we reject the null hypothesis that $\psi_{P_2}^{ne} = \psi_{P_2}^{rt}$, we also reject the composite null hypothesis that $\psi_{P_2}^{ne} = \psi_{P_2}^{rt}$ and $\psi_{P_3}^{ne} = \psi_{P_3}^{rt}$. Such a test is thus a falsification test
for the composite null hypothesis. In addition, since \[|\psi_{Z}^{ne} - \psi_{Z}^{rt}| \le  |\psi_{P_2}^{ne} - \psi_{P_2}^{rt}| + |\psi_{P_3}^{ne} - \psi_{P_3}^{rt}|\] by the triangle inequality,
$|\psi_{Z}^{ne} - \psi_{Z}^{rt}|$ also provides a natural lower bound for the sum of the absolute differences between $\psi_{P_2}^{ne}$ 
and $\psi_{P_2}^{rt}$ and between $\psi_{P_3}^{ne}$ and $\psi_{P_3}^{rt}$. Consequently, when $|\psi_{Z}^{ne} - \psi_{Z}^{rt}|$ is large, then large inconsistency (at least half as large as the bound) presents between either $\psi_{P_2}^{ne}$ and $\psi_{P_2}^{rt}$, or $\psi_{P_3}^{ne}$ and $\psi_{P_3}^{rt}$, or both. A toy example illustrating the informativeness of this bound is provided in Figure \ref{fig:2a}. 

\paragraph{Upper bounds of the path-specific differences} Deriving an upper bound for the absolute difference between recanting-twin and natural effects is also possible when the outcome is bounded. In particular, consider the case where the outcome $Y$ is binary, i.e., $Y \in \{0,1\}$. Proposition \ref{prop1} below establishes a nonparametric, identifiable upper bound for the absolute difference between recanting-twin and natural PSEs along paths $P_2$ and $P_3$. 

\begin{proposition} \label{prop1}
    For $Y\in \{0,1\}$, under the NPSEM-IE encoded in the causal diagram \ref{fig:1}, one has $$|\psi_{P_2}^{ne}-\psi_{P_2}^{rt}| \le \min (A,B) \qquad and \qquad |\psi_{P_3}^{ne}-\psi_{P_3}^{rt}| \le  \min(A_1,B_1)$$
    where:
    \begin{align*}
        A &= \max\big( |\P(Y_{S_1}=1) - 1 - \psi_{P_2}^{rt}|,|\P(Y_{S_1}=1)- \psi_{P_2}^{rt}|\big)\\[2pt]
        B &= 2 - \big[\max\{\P(Y_{S_1}=1), \P(Y_{S_1}'=0), \P(Y_{S_2}'=1)\}\\
    &\quad \quad - \min\{\P(Y_{S_1}=1), \P(Y_{S_1}'=0), \P(Y_{S_2}'=1)\}\big]\\
    &\quad \quad -\max\big\{ |\P(Y_{S_1}'=1)-\P(Y_{S_2}'=0)|, |\P(Y_{S_1}=1) -\P(Y_{S_1}'=0)|\big\}\\[2pt]
    A_1 &= \max\big( |\P(Y_{S_3}=1) + \psi_{P_3}^{rt}|,|1-\P(Y_{S_3}=1)- \psi_{P_3}^{rt}|\big)\\[2pt]
    B_1&=2 - \big[\max\{\P(Y_{S_3}=0), \P(Y_{S_2}''=0), \P(Y_{S_3}''=1)\}\\ 
    &\quad \quad- \min\{\P(Y_{S_3}=0), \P(Y_{S_2}''=0), \P(Y_{S_3}''=1)\}\big]\\
    &\quad \quad- \max\{ |\P(Y_{S_2}''=1)-\P(Y_{S_3}''=0)|, |\P(Y_{S_3}=1) -\P(Y_{S_3}''=0)|\}
    \end{align*}
\end{proposition}
Component $A$ and $A_1$ follow directly from the fact that the non-identifiable probability $\P(Y_{S_2}=1)$ is constrained to lie in $[0,1]$. By contrast, component $B$ and $B_1$ are obtained through a sharper argument. Specifically, applying Jensen’s inequality yields
$$|\psi_{P_2}^{ne}-\psi_{P_2}^{rt}| \le \E\{|Y_{S_1}-Y_{S_2}-Y_{S_1}'+Y_{S_2}'|\}$$
$$|\psi_{P_3}^{ne}-\psi_{P_3}^{rt}| \le \E\{|Y_{S_2}-Y_{S_3}-Y_{S_2}''+Y_{S_3}''|\}$$
The expectations on the right-hand sides are then further bounded using the structural relationships:
\[
Y_{S_1}=Y_{S_2} \iff Y_{S_1}'=Y_{S_2}',
\qquad\qquad
Y_{S_1}=Y_{S_1}' \iff Y_{S_2}=Y_{S_2}'\]
\[Y_{S_2}=Y_{S_3} \iff Y_{S_2}''=Y_{S_3}'',
\qquad\qquad
Y_{S_2}=Y_{S_2}'' \iff Y_{S_3}=Y_{S_3}''
\]
A detailed derivation is provided in Appendix B. A toy example illustrating the informativeness of these bounds is also provided in Figure \ref{fig:2b} and \ref{fig:2c}.
\subsection{Agreement in sign and ranking of natural and recanting-twin effects in non-linear settings: a simulation study}
We now conduct a simulation study to assess the concordance in sign and ranking between the two types of PSEs under settings that incorporate nonlinearity and non-monotonicity.  For simplicity, we assume that $Z$ and $M$ are univariate, and that no baseline covariate $W$ is present.  However, the findings generalize to settings with non-empty $W$ by focusing on conditional PSEs given $W$. 

Throughout the simulation, $Z$, $M$ and $Y$ can be either binary or continuous, which gives rise to eight distinct simulation scenarios. In each scenario, data are generated from a structural causal model (SCM) in which the potential variables $Z(a), M(a,z)$, and $Y(a,m,z)$ are simulated directly, bypassing explicit generation of the treatment assignment $A$. We evaluate the impact of four key parameters on the sign and ranking of the natural and recanting-twin PSEs. These parameters include (i) the direct effect of $A$ on $Z$, (ii) the direct effect of $Z$ on $M$, (iii) the direct effect of $Z$ on $Y$ and (iv) the correlation between $Z(0)$ and $Z(1)$. 

For each parameter configuration, we first compute the true values of the recanting-twin and natural PSEs via large-scale simulation with a sample size of $10^7$, using the true model coefficients. We then record the number of paths among $P_j $ for $ j=1,\ldots, 4$ whose relative rankings differ under the natural versus recanting-twin estimands. For example, if 
\begin{align*}
\psi_{P_1}^{ne}>\psi_{P_2}^{ne}>\psi_{P_3}^{ne}>\psi_{P_4}^{ne}\qquad \mathrm{and} \qquad 
\psi_{P_2}^{rt}>\psi_{P_3}^{rt}>\psi_{P_1}^{rt}>\psi_{P_4}^{rt}
\end{align*}
then the two systems of PSEs disagree in the ranking of three paths.

By definition, natural and recanting-twin effects are identical for paths $P_1$ and $P_4$, but not for paths $P_2$ and $P_3$. We thus also evaluate sign agreement between natural and recanting-twin PSEs for these two paths. Finally, we compute the magnitude of the recanting-twin parameter  $\psi_{P2 \vee P3}$ and assess whether it can predict inconsistencies between the two types of PSEs (see below).

In what follows, we describe the data-generating models and results for the four scenarios in which the mediator $M$ is continuous. The data generation and results of scenarios with  binary $M$ are provided in Appendix C (settings 3.1 to 4.2).

\paragraph{Scenario 1: Continuous outcome, continuous mediator}
For the case of continuous $Z$ (Scenario 1.1), data is generated by the following models:
\begin{equation}
\label{eq:mecha}
    \begin{aligned}
        \left(\begin{array}{c} Z(0) \\ Z(1) \end{array}\right)  &\sim \mathcal{N}\left[ \left( \begin{array}{c} \alpha_0 \\ \alpha_0+\alpha_1 \end{array} \right), \left(\begin{array}{cc}
            1 & \rho \\
             \rho & 1
        \end{array}\right)\right]\\
        M(a,z) &= \beta_0 + \beta_1 a + \beta_2 z + \beta_3 az + \varepsilon_M, \quad \varepsilon_M \sim \mathcal{N}(0, 1)\\
        Y(a,z,m) &= \gamma_0 + \gamma_1 a + \gamma_2 z + \gamma_3 m  + \gamma_4 az + \gamma_5 am + \gamma_6 zm + \gamma_7 azm + \varepsilon_Y, \quad \varepsilon_Y \sim \mathcal{N}(0, 1)\\
    \end{aligned}
\end{equation}
For the case of binary $Z$ (Scenario 1.2), we retain the same model for $Y$ and $M$, and generate $Z(0)$ and $Z(1)$ by a bivariate Bernoulli distribution, i.e.:
\begin{equation}
    \label{eq:Zbi}
    \begin{aligned}
        Z(a) &\sim \text{Bern}(\text{expit}(\alpha_0 + \alpha_1 a))\\
        \mathrm{corr}&\{Z(0),Z(1)\} =\rho
    \end{aligned}
\end{equation}

In both settings 1.1 and 1.2, the key parameters $\rho$, $\alpha_1$, $\beta_2$ and $\gamma_6$ can take different values as follow:
\begin{align*}
    \rho &= \begin{pmatrix} -0.75& -0.20& 0.20& 0.75 \end{pmatrix}\\
    \alpha_1 &= \begin{pmatrix} -1.50& -0.50& 0.50& 1.50 \end{pmatrix}\\
    \beta_2 &=\begin{pmatrix} -1.50& -1.00& 1.00& 1.50 \end{pmatrix}\\
    \gamma_6 &= \begin{pmatrix} -1.50& -1.00& 0.00 & 1.00& 1.50 \end{pmatrix}
\end{align*}
In contrast, all other parameters are fixed at prespecified values (see Appendix C for a detailed description).
Finally, it is worth noting that the recanting-twin variables $T(a)$ for $a=0,1$ are drawn independently from the marginal distribution of $Z(a)$. More specifically, in setting 1.1 (continuous $Z$):
    \begin{align*}
        T(0) \sim \mathcal{N}(\alpha_0,1)  \qquad \mathrm{and} \qquad T(1) \sim \mathcal{N}(\alpha_0+\alpha_1,1) 
    \end{align*}
whilst in setting 1.2 (binary $Z$):
    \begin{align*}
        T(0) \sim \text{Bern}(\alpha_0) \qquad \mathrm{and} \qquad T(1) \sim \text{Bern}(\alpha_0+\alpha_1) 
    \end{align*}
\paragraph{Scenario 2: Binary outcome, continuous mediator}
We now consider the outcome $Y$ to be binary, while $Z$ is either continuous (Scenario 2.1) or binary (Scenario 2.2). The data-generating mechanism of $Z$ and $M$ stays the same as in Scenario 1, with continuous $Z$ generated by model \ref{eq:mecha} and binary $Z$ generated by model \ref{eq:Zbi}. In contrast, the binary outcome $Y$ is generated as:
\begin{align*}
    Y(a,m,z) \sim \text{Bern}(\text{expit}(\gamma_0 + \gamma_1a + \gamma_2z+ \gamma_3m + \gamma_4az+ \gamma_5am + \gamma_6zm + \gamma_7azm)) 
\end{align*}

\noindent The set of values for parameters ($\rho$, $\alpha_1$, $\beta_2$, $\gamma_6$) remains the same as in Scenario 1. 


\paragraph{Simulation results}

Results of this simulation studies are visualized in Figures \ref{fig:3}, \ref{fig:4}, and Appendix C (Figures \ref{fig:a6} and \ref{fig:a7}). These figures display (i) the number of paths for which the ranking of effects differs between the natural and recanting-twin estimands (first row),  and (ii) the agreement in sign of path $P_2$ (second row) and $P_3$ (third row), with the size of the dots representing the magnitude of $\psi_{P2 \vee P3}$. 
\\\\
\noindent \textbf{Ranking} -- Natural and recanting-twin systems yield inconsistent rankings of the path-specific effects in 37.5\% of settings under Scenario 1.1 (continuous $Z$, $M$, $Y$), 59.7\% under Scenario 1.2 (binary $Z$, continuous $M$ and $Y$), 47.8\% under Scenario 2.1 (continuous $Z$ and $M$, binary $Y$) and 56.2\% of settings in Scenario 2.2 (binary $Z$, continuous $M$ and binary $Y$). Ranking inconsistencies are most often driven by disagreements in two or three paths out of four, although complete disagreement across all four paths is also observed. A consistent pattern across all scenarios is that ranking discrepancies tend to become more frequent as the correlation $\rho$ between $Z(0)$ and $Z(1)$ becomes increasingly negative. In contrast, discrepancy may also more often arise from specific combinations of other model parameters $\alpha_1$, $\beta_2$ or $\gamma_6$, but no consistent patterns were observed. In particular, the parameter $\psi_{P_2 \vee P_3}$, proposed as a measure of the severity of intermediate confounding induced by $Z$, does not appear to be associated with the extent of ranking disagreement in any of the simulated scenarios.     
\\\\
\noindent \textbf{Sign} -- Natural and recanting-twin systems yield PSE estimands with discordant sign for path $P_2$ ($A \rightarrow Z \rightarrow Y$) in 8.1\% of settings under Scenario 1.1; 13.1\% under Scenario 1.2; 27.5\% under Scenario 2.1; and 24.1\% under Scenario 2.2. The frequency of sign disagreement is slightly higher for path $P_3$ ($A \rightarrow Z \rightarrow M \rightarrow Y$), reaching 18.8\%, 23.3\%, 18.4\% and 36.7\% in Scenario 1.1, 1.2, 2.1 and 2.2, respectively. 

To further characterize these discrepancies, we examine within each scenario the magnitude of the absolute difference $|\psi_j^{ne}-\psi_j^{rt}|$ when $\psi_j^{ne}$ and $\psi_j^{rt}$ have different sign, for $j=2,3$ (Appendix B). For a given outcome and mediator type, this absolute difference is substantially smaller when $Z$ is binary (Scenario 1.2 and 2.2) than when $Z$ is continuous (Scenario 1.1 and 2.1). Notably, in Scenario 2.2 (binary $Z$, continuous $M$ and binary $Y$), although sign disagreement between $\psi_j^{ne}$ and $\psi_j^{rt}$ for $j=2,3$ occurs relatively frequently, the corresponding absolute differences remain small, typically below 0.05.

Finally, across all scenarios, the prevalence of sign disagreement appears to decrease with positively
increased $\rho$. Sign disagreement also tends to cluster around specific combinations of $\alpha_1$, $\beta_2$ and $\gamma_6$. In contrast, the magnitude of $\psi_{P2 \vee P3}$ does not appear to predict the occurrence of sign inconsistency.

\section{Discussion}

The recanting twin approach proposed by \cite{diaz2022causal} addresses the issue of intermediate confounding in causal mediation analysis, enabling the decomposition of the average treatment effect into identifiable path-specific effects. An open question for this approach, however, is the extent to which PSEs defined under the recanting-twin approach can inform the direction and relative magnitude of the corresponding natural PSEs. 

In this paper, we systematically characterize the behaviors of recanting-twin PSEs in terms of magnitude, sign and ranking, benchmarking against natural PSEs. We show that in the presence of intermediate confounder, the PSEs under two systems only exactly coincides under linear or log-linear outcome models when there is no interactions between the mediator and the intermediate confounder, or when additional strong assumptions on the data generating process hold. In more general settings, the PSEs under two systems often diverge. Nevertheless, when the outcome of interest is binary, we derive non-parametric, identifiable upper and lower bounds for the absolute difference between recanting-twin and natural PSEs. These bounds provide a practical way to assess whether discrepancies in magnitude between the two types of effects are substantial, thereby facilitating the use of recanting-twin PSEs as informative approximations of their natural counterparts. 

One central challenge, however, is that when individual-level PSEs vary in opposite directions across subjects in the target population, disagreements in sign and ranking between recanting-twin and natural effects become relatively common at the population level. Across 320 evaluated simulation settings, we find that disagreement ranged from 15\% to 56\% for ranking and from 5\% to 43\% for sign. The occurrence of such disagreement is strongly influenced by the correlation between the counterfactual values of the intermediate confounder, as well as by the complex interplay among other parameters governing the data-generating process. These findings suggest that relying exclusively on the recanting-twin approach to quantify and compare path-specific effects remains challenging, particularly because some of the parameters driving these inconsistencies cannot be identified or learned in practice (e.g., the correlation between counterfactual values of the intermediate confounder). 
Further research is therefore needed to improve recanting-twin PSEs or, at a minimum, to identify conditions under which recanting-twin and natural PSE systems yield consistent conclusions regarding effect sign and path ranking.

This work has some limitations. First, in our simulation study, parameter values were selected without reference to a specific clinical or policy context, which prevents us from assessing how inconsistencies in sign and ranking would translate into differences in substantive conclusions in applied settings. Second, although the proposed bounds for the absolute difference between natural and recanting-twin PSEs are fully non-parametric, and therefore robust to model misspecification, they are not smooth or pathwise differentiable. As a result, establishing their asymptotic properties using standard non-parametric theory becomes challenging, particularly when data-adaptive estimation methods are employed. By contrast, when parametric estimation approaches are used, uncertainty associated with these bounds can be quantified using bootstrap procedures.

\section*{Funding}
T.T.V is supported by the French National Research Agency (Agence Nationale de la Recherche), through a funding for Chaires de Professeur Junior (23R09551S-MEDIATION).
\section*{Conflict of interest}
All authors declare that they have no conflicts of interest.

\bibliographystyle{plainnat}
\bibliography{reference}
\newpage
\section*{Tables and Figures}
\newpage
\begin{landscape}
    \begin{figure}[H]
        \centering
        \includegraphics[width=0.68\linewidth]{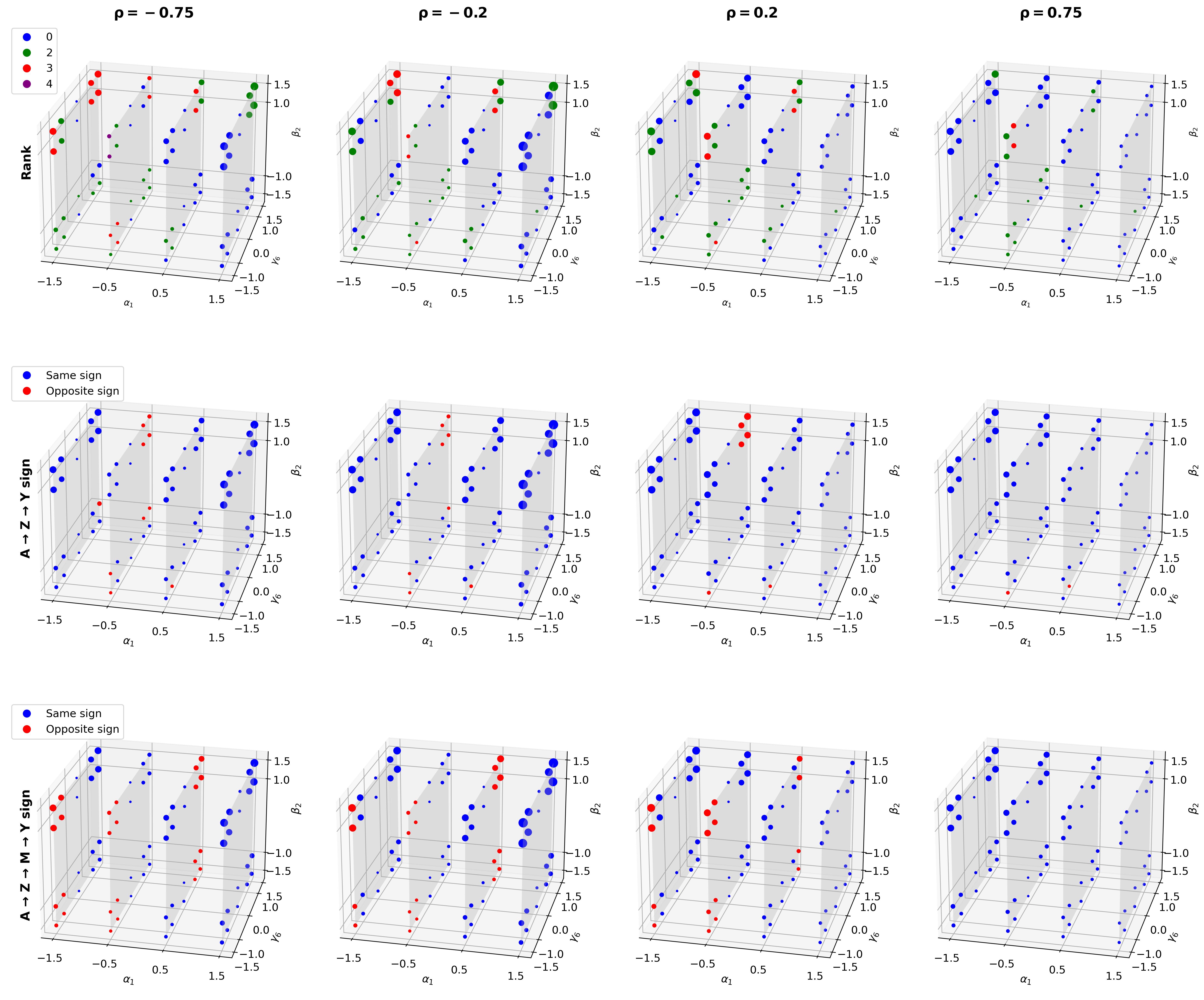}
        \caption{Disagreement in ranking and sign of natural and recanting-twin PSEs in  setting 1.1 (continuous $Z$, $M$, $Y$). First panel: number of paths among $P_j:j=1,\ldots,4$ whose relative ranking differ under the natural versus recanting-twin system. Second panel: red and blue dots represent settings with disagreement and agreement in sign between $\psi_{P_2}^{ne}$ and $\psi_{P_2}^{rt}$, respectively. Third panel: red and blue dots represent settings with disagreement and agreement in sign between $\psi_{P_3}^{ne}$ and $\psi_{P_3}^{rt}$, respectively. The size of the dots is relative to the magnitude of the intermediate confounding parameter $\psi_{P2 \vee P3}$.}
        \label{fig:3}
    \end{figure}
    \newpage
    \begin{figure}[H]
        \centering
        \includegraphics[width=0.68\linewidth]{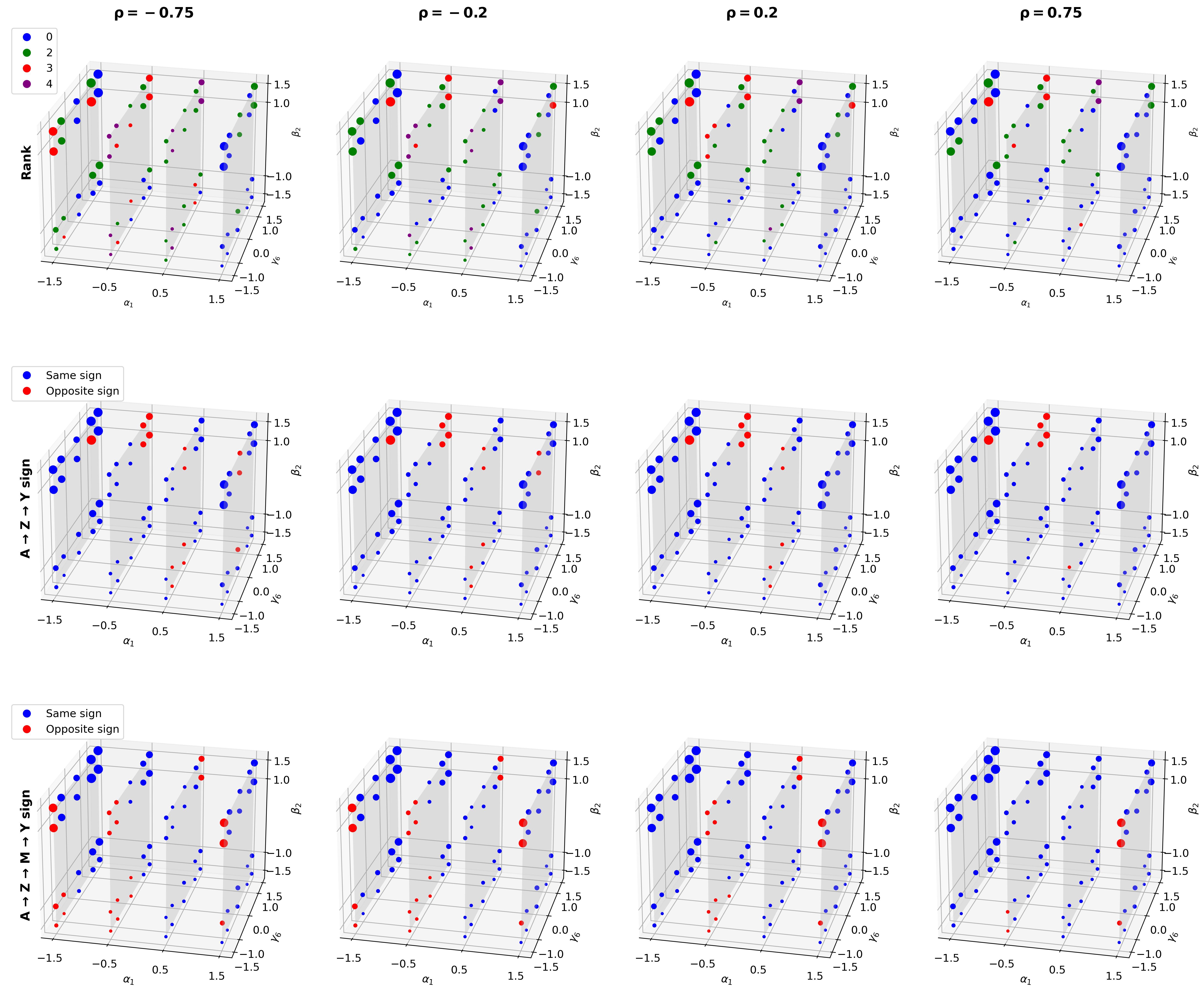}
        \caption{Disagreement in ranking and sign of natural and recanting-twin PSEs in  setting 2.1 (continuous $Z$ and $M$, binary $Y$). First panel: number of paths among $P_j:j=1,\ldots,4$ whose relative ranking differ under the natural versus recanting-twin system. Second panel: red and blue dots represent settings with disagreement and agreement in sign between $\psi_{P_2}^{ne}$ and $\psi_{P_2}^{rt}$, respectively. Third panel: red and blue dots represent settings with disagreement and agreement in sign between $\psi_{P_3}^{ne}$ and $\psi_{P_3}^{rt}$, respectively. The size of the dots is relative to the magnitude of the intermediate confounding parameter $\psi_{P2 \vee P3}$.}
        \label{fig:4}
    \end{figure}
\end{landscape}
\newpage 
\section*{Appendix}
\appendix
\renewcommand{\thetable}{A\arabic{table}}
\renewcommand{\thefigure}{A\arabic{figure}}
\section{Closed-form expressions of $\mathbf{\psi^{ne}_{P_2}}$, $\mathbf{\psi^{re}_{P_2}}$, $\mathbf{\psi^{ne}_{P_3}}$, $\mathbf{\psi^{re}_{P_3}}$ under linear outcome model}

Consider the following linear outcome model:
\begin{align*}
\E[Y(a,z,m) \mid W] = \gamma_0^{a,W} + \gamma_{1}^{a,W}z + \gamma_{2}^{a,W}m + \gamma_{3}^{a,W}zm
\end{align*}

\subsubsection*{Identification of $\mathbf{\E(Y_{S_j}|W)}$ 
$\mathbf{(j=\overline{1,3})}$, $\psi^{ne}_{P_2}$ and $\psi^{ne}_{P_3}$}
Under the above outcome model:
\begin{align*}
& \E[Y(a_1,Z(a_2),M(a_3,Z(a_4))|W] \\
& = \sum_{z,m}\E(Y|a_1,z,m,W) \times \P\big(M(a_3,Z(a_4))=m|a_1,z,W\big) \times \P(Z(a_2)=z|a_1,W) \\
& = \sum_{z,m}(\gamma^{a_1,W}_0+\gamma^{a_1,W}_1z + \gamma^{a_1,W}_2m + \gamma^{a_1,W}_3zm) \times \P(M(a_3,Z(a_4))=m|a_1,z,W) \times P(Z(a_2)=z|a_1,W) \\
& = \gamma^{a_1,W}_0 + \gamma^{a_1,W}_1\E(Z(a_2)|a_1,W) + \gamma^{a_1,W}_2\E(M(a_3,Z(a_4))|a_1,W) + \gamma^{a_1,W}_3\E(Z(a_2)M(a_3,Z(a_4))|a_1,W)
\end{align*}

\noindent From this general formula, we then have:
\begin{align*}
& \E(Y_{S_1}) = \E[Y(a^{*}, Z(a), M(a,Z(a))] \text{ (with } a_1=a^*, a_2=a_3=a_4=a) \\
& = \gamma^{a^*,W}_0 + \gamma^{a^*,W}_1\E(Z(a)|a^*,W) + \gamma^{a^*,W}_2\E(M(a,Z(a))|a^*,W) + \gamma^{a^*,W}_3\E(Z(a)M(a,Z(a))|a^*,W) \\
& = \gamma^{a^*,W}_0 + \gamma^{a^*,W}_1\E(Z|a,W) + \gamma^{a^*,W}_2\E(M|a,W) + \gamma^{a^*,W}_3\E(Z(a)M(a,Z(a))|W) \\
&\quad \text{ (as } Z(a), M(a,Z(a)) \perp\!\!\!\!\perp A|W) \\
\\
& \E(Y_{S_2}) = \E(Y[a^{*},Z(a^{*}),M(a,Z(a)] \text{ (with } a_1=a_2=a^{*},a_3=a_4=a) \\ 
& = \gamma^{a^*,W}_0 + \gamma^{a^*,W}_1\E(Z(a^*)|a^*,W) + \gamma^{a^*,W}_2\E(M(a,Z(a))|a^*,W) + \gamma^{a^*,W}_3\E(Z(a^*)M(a,Z(a))|a^*,W) \\
& = \gamma^{a^*,W}_0 + \gamma^{a^*,W}_1\E(Z|a^*,W) + \gamma^{a^*,W}_2\E(M|a,W) + \gamma^{a^*,W}_3\E(Z(a^*)M(a,Z(a))|W) \\
&\quad \text{ (as } Z(a^*), M(a,Z(a)) \perp\!\!\!\!\perp A|W) \\
\\
& \E(Y_{S_3}) = \E(Y(a^*,Z(a^*),M(a,Z(a^*)) \text{ (with } a_1=a_2=a_4=a^{*},a_3=a) \\
& = \gamma^{a^*,W}_0 + \gamma^{a^*,W}_1\E(Z(a^*)|a^*,W) + \gamma^{a^*,W}_2\E(M(a,Z(a^*))|a^*,W) + \gamma^{a^*,W}_3\E(Z(a^*)M(a,Z(a^*))|a^*,W) \\
& = \gamma^{a^*,W}_0 + \gamma^{a^*,W}_1\E(Z|a^*,W) + \gamma^{a^*,W}_2\E(M(a,Z(a^*))|a^*,W) + \gamma^{a^*,W}_3\E(Z(a^*)M(a,Z(a^*))|W) \\
\end{align*}

\noindent The natural PSE $\psi_{P_2}^{ne}$ and $\psi_{P_3}^{ne}$ can therefore be expressed as:
\begin{align*}
 \psi_{P_2}^{ne} &= \E(Y_{S_1} - Y_{S_2} \mid W) \\
&= \gamma^{a^*,W}_1~\mathsf{ATE}^W_Z + \gamma^{a^*,W}_3\E(Z(a)M(a,Z(a))-Z(a^*)M(a,Z(a))|W) \\
&= \gamma^{a^*,W}_1~\mathsf{ATE}^W_Z + \gamma^{a^*,W}_3\E\{Z(a)M(a)-Z(a^*)M(a)|W\}. \blacksquare \\
\\
 \psi^{ne}_{P_3} &= \E(Y_{S_2} - Y_{S_3}\mid W) \\
&= \gamma^{a^*,W}_2[\E(M|a,W)-\E(M(a,Z(a^*))|a^*,W)] \\
&\quad + \gamma^{a^*,W}_3[\E(Z(a^*)M(a,Z(a))|W) - \E(Z(a^*)M(a,Z(a^*))|W)] \\
&= \gamma^{a^*,W}_2[\E(M(a)|W)-\E(M(a,Z(a^*))|W] \\
&\quad + \gamma^{a^*,W}_3[\E(Z(a^*)M(a)|W)- \sum_{z}z\E(M(a,z)|W)\P(Z(a^*)=z|W)] \\
&= \gamma^{a^*,W}_2~\mathsf{NIE}^W_M + \gamma^{a^*,W}_3[\E(Z(a^*,M(a)|W)-\sum_{z}z\E(M(a,z)|W)\P(Z(a^*)=z|W)]. \blacksquare
\end{align*}
where $~\mathsf{ATE}^W_Z=\E[Z(a)-Z(a^*)|W]$ is the average effect of $A$ on $Z$ given $W$, and $~\mathsf{NIE}^W_M=\E(M(a)-M(a,Z(a^*))|W)$ is the natural indirect effect of $A$ on $M$ via $Z$ given $W$.

\subsubsection*{Identification of $\mathbf{\E(Y'_{S_j})}$ and $\mathbf{\psi^{rt}_{P_2}}$}
Note that:
\begin{equation}
\label{eq:rt_s1s2} 
\begin{aligned}
&\E\left\{Y\left(a_1, Z(a_2), M(a_3, T(a_4))\right) | W \right\} \\
& = \sum_{z, z', m} \E\{Y(a_1, z, m) | Z(a_2) = z, T(a_4) = z', M(a_3, z') = m, W\} \\
&\quad \times \P(M(a_3, z') = m | T(a_4) = z', Z(a_2) = z, W) 
\times \P(T(a_4) = z' | Z(a_2) = z, W) \times \P(Z(a_2) = z | W) \\
&= \sum_{z, z', m} \E\{Y(a_1, z, m) | W\}
\times \P(M(a_3, z') = m | W)
\times \P(T(a_4) = z' | W)
\times \P(Z(a_2) = z | W) \\
&= \sum_{z, z', m} \E(Y | a_1, z, m, W) 
\times \P(M = m | a_3, z', W) 
\times \P(Z = z' | a_4, W) 
\times \P(Z = z | a_2, W)
\end{aligned}
\end{equation}

\noindent For $Y_{S1}'$, replace $a_1 = a^*, a_2 = a_3 = a_4 = a $ in equation \ref{eq:rt_s1s2}:
\begin{align*}
& \E(Y_{S_1}'|W) = \E(Y(a^*,Z(a),M(a,T(a))) \\ 
&= \sum_{z, z', m} \E(Y | a^*, z, m, W) 
    \times \P(M = m | a, z', W) 
    \times \P(Z = z' | a, W) 
    \times \P(Z = z | a, W) \\
&= \sum_{z,m}\{\E(Y | a^*, z, m, W)\times \P(Z = z | a, W) \times [\sum_{z'} \P(M = m | a, z', W) \times \P(Z = z' | a, W)]\} \\
&= \sum_{z,m}\E(Y | a^*, z, m, W)\times \P(Z = z | a, W) \times \P(M=m | a, W) \\
&= \sum_{z,m}(\gamma^{a^*,W}_0 + \gamma^{a^*,W}_1z + \gamma^{a^*,W}_2m + \gamma^{a^*,W}_3zm) \times \P(Z=z|a,W) \times \P(M=m|a,W) \\
&= \gamma^{a^*,W}_0 + \gamma^{a^*,W}_1\E(Z|a,W) + \gamma^{a^*,W}_2\E(M|a,W) + \gamma^{a^*,W}_3\E(Z|a,W)\E(M|a,W)
\end{align*}

\noindent For $Y_{S2}'$, replace $a_1 = a_2 = a^*,  a_3 = a_4 = a $ in equation \ref{eq:rt_s1s2}:
\begin{align*}
&\E(Y_{S_2}'|W) = \E(Y(a^*,Z(a^*),M(a,T(a)) \mid W) \\
&= \sum_{z, z', m} \E(Y | a^*, z, m, W) 
    \times \P(M = m | a, z', W) 
    \times \P(Z = z' | a, W) 
    \times \P(Z = z | a^*, W) \\
&= \sum_{z,m}\{\E(Y | a^*, z, m, W)\times \P(Z = z | a*, W) \times [\sum_{z'} \P(M = m | a, z', W) \times \P(Z = z' | a, W)] \} \\
&= \sum_{z,m}\E(Y | a^*, z, m, W)\times \P(Z = z | a^*, W) \times \P(M=m | a, W) \\
&= \sum_{z,m}(\gamma^{a^*,W}_0 + \gamma^{a^*,W}_1z + \gamma^{a^*,W}_2m + \gamma^{a^*,W}_3zm) \times \P(Z=z|a^*,W) \times \P(M=m|a,W) \\
&= \gamma^{a^*,W}_0 + \gamma^{a^*,W}_1\E(Z|a^*,W) + \gamma^{a^*,W}_2\E(M|a,W) + \gamma^{a^*,W}_3\E(Z|a^*,W)\E(M|a,W)
\end{align*}

\noindent Derive $\psi^{rt}_{P_2}$ from $Y'_{S_1}$ and $Y'_{S_2}$, we get:
\begin{align*}
\psi^{rt}_{P_2} = \E(Y'_{S_1}\mid W) - \E(Y'_{S_2}\mid W) =\gamma^{a^*,W}_1~\mathsf{ATE}^W_{Z} + \gamma^{a^*,W}_3\E(M|a,W)~\mathsf{ATE}^W_{Z}. \blacksquare
\end{align*}

\subsubsection*{Identification of $\mathbf{\E(Y''_{S_j})}$ and $\mathbf{\psi^{rt}_{P_3}}$}

\begin{equation}
\label{eq:rt_s2s3}
\begin{aligned}
&\E\left\{Y\left(a_1, T(a_2), M(a_3, Z(a_4))\right) \mid W \right\} \\
&= \sum_{z,z',m} \E\{Y(a_1, z, m) \mid T(a_2)=z, Z(a_4)=z', M(a_3, z')=m, W\} \\
    &\quad \times \P(M(a_3, z') = m \mid Z(a_4) = z', T(a_2) = z, W)
    \times \P(Z(a_4) = z' \mid T(a_2) = z, W)
    \times \P(T(a_2) = z \mid W) \\
&= \sum_{z,z',m} \E\{Y(a_1, z, m) \mid W\} \\
    &\quad \times \P(M(a_3, z') = m \mid W) \times \P(Z(a_4) = z' \mid W)
    \times \P(T(a_2) = z \mid W) \\
&= \sum_{z, z', m} \E(Y \mid a_1, z, m, W) 
    \times \P(M = m \mid a_3, z', W)
    \times \P(Z = z' \mid a_4, W)
    \times \P(Z = z \mid a_2, W)
\end{aligned}
\end{equation}

\noindent As \ref{eq:rt_s2s3} has the same formula as \ref{eq:rt_s1s2}, for $Y_{S2}''$, we get the same quantity as $Y_{S2}'$ by replacing $a_1 = a_2 = a^*,  a_3 = a_4 = a$ in equation \ref{eq:rt_s2s3}.
\\\\
For $Y_{S3}''$, replace $a_1 = a_2 = a_4 = a^*,  a_3 = a$ in equation \ref{eq:rt_s2s3}:
\begin{align*}
&\E(Y_{S_3}''|W) = \E(Y(a^*,T(a^*),M(a,Z(a^*))\mid W) \\
&= \sum_{z, z', m} \E(Y \mid a^*, z, m, W) 
    \times \P(M = m \mid a, z', W) 
    \times \P(Z = z' \mid a^*, W) 
    \times \P(Z = z \mid a^*, W) \\
&= \sum_{z,z',m}(\gamma^{a^*,W}_0 + \gamma^{a^*,W}_1z + \gamma^{a^*,W}_2m + \gamma^{a^*,W}_3zm) \\
&\quad \times \P(M = m \mid a, z', W) 
    \times \P(Z = z' \mid a^*, W) 
    \times \P(Z = z \mid a^*, W) \\
&= \gamma^{a^*,W}_0 + \gamma^{a^*,W}_1\E(Z|a^*,W) + \gamma^{a^*,W}_2\E(M|a,z',W)\P(Z=z'|a^*,W) \\
&\quad+ \gamma^{a^*,W}_3\E(Z|a^*,W)\E(M|a,z',W)\P(Z=z'|a^*,W) \\
&= \gamma^{a^*,W}_0 + \gamma^{a^*,W}_1\E(Z|a^*,W) + \gamma^{a^*,W}_2\E(M(a)|a^*,W) + \gamma^{a^*,W}_3\E(Z|a^*,W)\E(M(a)|a^*,W)
\end{align*}
\\\\
\noindent Derive $\psi^{rt}_{P_3}$ from $Y''_{S_2}$ and $Y''_{S_3}$, we get:
\begin{align*}
&\psi^{rt}_{P_3} = \E(Y''_{S_2}) - \E(Y''_{S_3}) = \gamma^{a^*,W}_2~\mathsf{NIE}^W_M + \gamma^{a^*,W}_3\E(Z|a^*,W)~\mathsf{NIE}^W_M. \blacksquare
\end{align*}

\section{Bounding the absolute differences between natural and recanting-twin PSEs in binary outcome settings}
By Jensen's inequality:
\begin{align*}
    |\psi_{P_2}^{ne}-\psi_{P_2}^{rt}| &= |\E(Y_{S_1}-Y_{S_2})-\E(Y_{S_1}'-Y_{S_2}')|\\
    &\le \E\{|Y_{S_1}-Y_{S_2}-Y_{S_1}'+Y_{S_2}'|\}
\end{align*}
Now note that:
\[
 |Y_{S_1}-Y_{S_2}-Y_{S_1}'+Y_{S_2}'|
 = \begin{cases}
    2 & \text{if } (Y_{S_1}=1, Y_{S_2}=Y_{S_1}'=0,Y_{S_2}'=1) \text{ or } (Y_{S_1}=0, Y_{S_2}=Y_{S_1}'=1,Y_{S_2}'=0)\\
    0 & \text{if } Y_{S_1}-Y_{S_2}=Y_{S_1}'-Y_{S_2}'\\
    1 & \text{otherwise}
\end{cases}
\] 
Hence,
\begin{align*}
  &\E(|Y_{S_1}-Y_{S_2}-Y_{S_1}'+Y_{S_2}'|)\\
  &=2\times\P(A) + 1\times[1-\P(A)-\P(Y_{S_1}-Y_{S_2}=Y_{S_1}'-Y_{S_2}']\\
  &=1+\P(A)-\P(Y_{S_1}-Y_{S_2}=Y_{S_1}'-Y_{S_2}')
\end{align*}
where: 
\begin{align*}
    \P(A)&=\P(Y_{S_1}=1, Y_{S_2}=Y_{S_1}'=0,Y_{S_2}'=1) + \P(Y_{S_1}=0, Y_{S_2}=Y_{S_1}'=1,Y_{S_2}'=0)\\
    &\le \min\{\P(Y_{S_1}=1),\P(Y_{S_1}'=0),\P(Y_{S_2}'=1)\} + \min \{\P(Y_{S_1}=0),\P(Y_{S_1}'=1),\P(Y_{S_2}'=0)\}\\
    &=\min\{\P(Y_{S_1}=1),\P(Y_{S_1}'=0),\P(Y_{S_2}'=1)\} + 1 - \max\{\P(Y_{S_1}=1),\P(Y_{S_1}'=0),\P(Y_{S_2}'=1)\} 
\end{align*}
Here, the last line results from the fact that $\min(1-x,1-y,1-z)=1-\max(x,y,z)$.

Now notice that:
\[\P(Y_{S_1}-Y_{S_2}=Y_{S_1}'-Y_{S_2}') \ge \P(Y_{S_1}-Y_{S_2}=Y_{S_1}'-Y_{S_2}'=0) = \P(Y_{S_1}'-Y_{S_2}'=0)\]
The last equality results from the fact that $\{Y_{S_1}=Y_{S_2}\} = \{Y_{S_1}'=Y_{S_2}'\}$. In what follows:
\begin{align*}
    \P(Y_{S_1}'=Y_{S_2}') &=1 - \P(Y_{S_1}'\ne Y_{S_2}')\\
    &=1-[\P(Y_{S_1}'=1, Y_{S_2}'=0)+\P(Y_{S_1}'=0, Y_{S_2}'=1)]\\
    &\ge 1-[\min\{\P(Y_{S_1}'=1), \P(Y_{S_2}'=0)\} + \min\{\P(Y_{S_1}'=0), \P(Y_{S_2}'=1)\}]\\
    &=1-[\min\{\P(Y_{S_1}'=1), \P(Y_{S_2}'=0)\} + 1 - \max\{\P(Y_{S_1}'=1), \P(Y_{S_2}'=0)\}]\\
    &=\max\{\P(Y_{S_1}'=1), \P(Y_{S_2}'=0)\}- \min\{\P(Y_{S_1}'=1), \P(Y_{S_2}'=0)\}\\
    &=|\P(Y_{S_1}'=1)-\P(Y_{S_2}'=0)|
\end{align*}
where the fourth line results from the fact that $\min(1-x,1-y)=1-\max(x,y)$. 
\\\\
We also have that:
\begin{align*}
    \P(Y_{S_1}-Y_{S_2}=Y_{S_1}'-Y_{S_2}') & = \P(Y_{S_1}-Y_{S_1}'=Y_{S_2}-Y_{S_2}')\\
    &\ge \P(Y_{S_1}-Y_{S_1}'=Y_{S_2}-Y_{S_2}'=0) \\
    &= \P(Y_{S_1}-Y_{S_1}'=0)\\
    &\ge |\P(Y_{S_1}=1) -\P(Y_{S_1}'=0)|
\end{align*}
which implies that:
\begin{align*}
    \P(Y_{S_1}-Y_{S_2}=Y_{S_1}'-Y_{S_2}') &\ge {\max\{ |\P(Y_{S_1}'=1)-\P(Y_{S_2}'=0)|, |\P(Y_{S_1}=1) -\P(Y_{S_1}'=0)|\}}
\end{align*}
We obtain the following bound:
\begin{align*}
    |\psi_{P_2}^{ne}-\psi_{P_2}^{rt}| \le 2 &- \big[\max\{\P(Y_{S_1}=1), \P(Y_{S_1}'=0), \P(Y_{S_2}'=1)\}\\ 
    &- \min\{\P(Y_{S_1}=1), \P(Y_{S_1}'=0), \P(Y_{S_2}'=1)\}\big]\\
    &- \max\{ |\P(Y_{S_1}'=1)-\P(Y_{S_2}'=0)|, |\P(Y_{S_1}=1) -\P(Y_{S_1}'=0)|\}
\end{align*}
Similarly, 
\begin{align*}
    |\psi_{P_3}^{ne}-\psi_{P_3}^{rt}| \le 2 &- \big[\max\{\P(Y_{S_3}=0), \P(Y_{S_2}''=0), \P(Y_{S_3}''=1)\}\\ 
    &- \min\{\P(Y_{S_3}=0), \P(Y_{S_2}''=0), \P(Y_{S_3}''=1)\}\big]\\
    &- \max\{ |\P(Y_{S_2}''=1)-\P(Y_{S_3}''=0)|, |\P(Y_{S_3}=1) -\P(Y_{S_3}''=0)|\}
\end{align*}

To illustrate the optimal bounds of the proposed method, we conduct a simulation study under the following parameter settings. We set $\rho=-0.2$ and consider the following parameter grids:
\begin{align*}
    \alpha_1 &= \begin{pmatrix} -1.5&-1.0&-0.5&0.0&0.5&1.0&1.5 \end{pmatrix}\\
    \beta_2 &=\begin{pmatrix} -1.5&-1.0&-0.5&0.0&0.5&1.0&1.5 \end{pmatrix}\\
    \gamma_2 &= \begin{pmatrix} -1.5&-1.0&-0.5&0.0&0.5&1.0&1.5 \end{pmatrix}\\
    \gamma_3 &= \begin{pmatrix} -1.5&-1.0&-0.5&0.0&0.5&1.0&1.5 \end{pmatrix}\\
    \gamma_6 &= \begin{pmatrix} -1.5&-1.0&-0.5&0.0&0.5&1.0&1.5 \end{pmatrix}
\end{align*}
We consider four scenarios based on the types of $Z$ and $M$: (i) $Z$ continuous and $M$ continuous, (ii) $Z$ continuous and $M$ binary, (iii) $Z$ binary and $M$ continuous, and (iv) $Z$ binary and $M$ binary. The outcome variable $Y$ is generated as a binary variable according to the specified data-generating mechanism. For each scenario, we investigate the sensitivity of the upper bounds of $|\psi_{P_2}^{ne} - \psi_{P_2}^{rt}|$ and $|\psi_{P_3}^{ne} - \psi_{P_3}^{rt}|$, as well as the lower bound of $|\psi_{P_2}^{ne} - \psi_{P_2}^{rt}|$ + $|\psi_{P_3}^{ne} - \psi_{P_3}^{rt}|$, to the parameters of the data-generating mechanism. We select the parameter that yields the most pronounced variation in the corresponding bound and use this parameter configuration for illustration purposes only. 

\newpage
\section{Details of the simulation study}

\begin{figure}[H]
    \centering
    \includegraphics[width=0.9\linewidth]{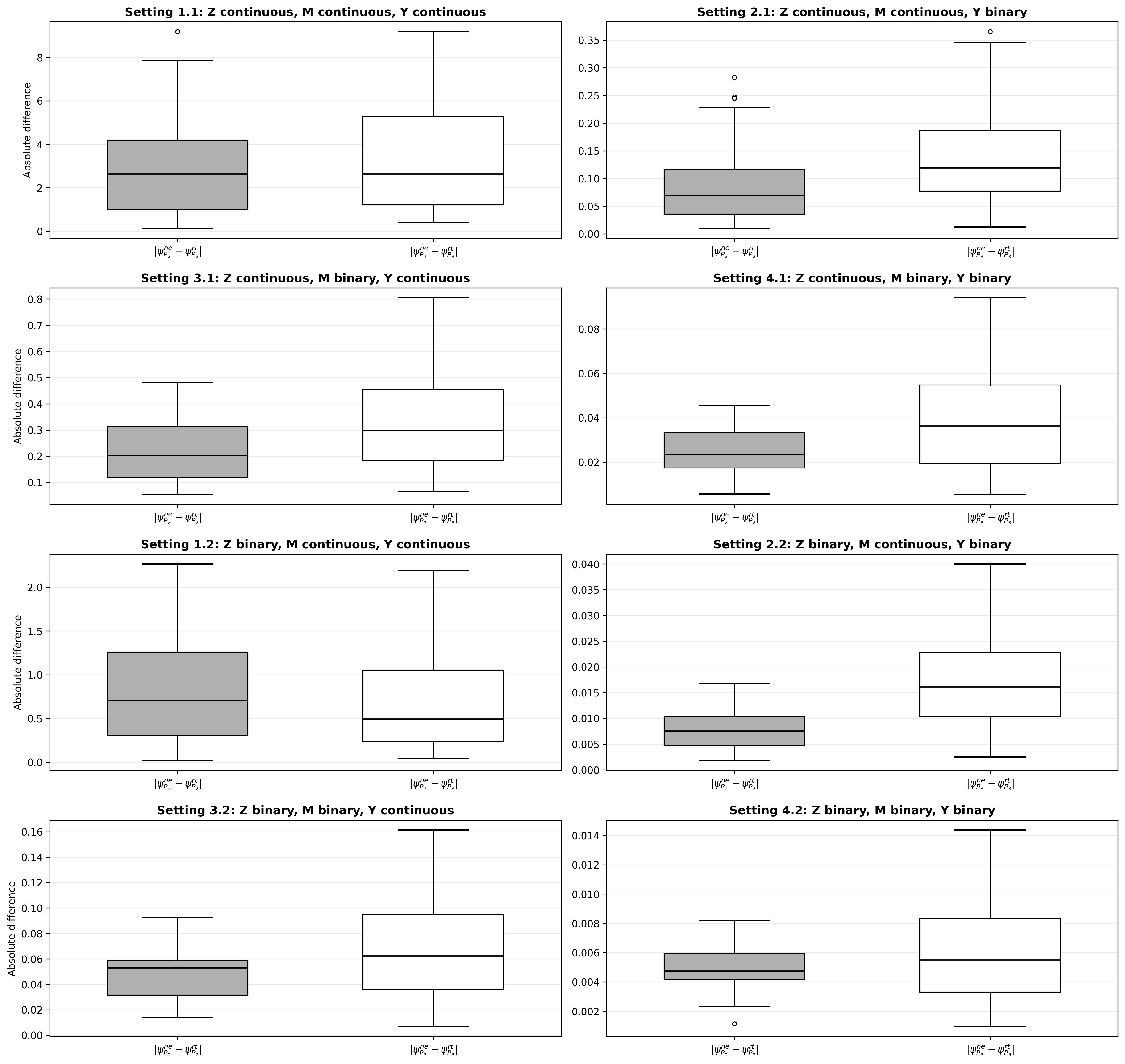}
    \caption{Distribution of the absolute difference between natural and recanting-twin effects when they have opposite signs across all settings.}
    \label{fig:a5}
\end{figure}

Eight scenarios corresponding to all combinations of $Z$, $M$, and $Y$ being continuous or binary are encoded as followed.
\begin{center}
    \begin{table}[H]
        \centering
        \caption{Encoding of the eight simulation scenarios}
        \label{tab:code}
        \renewcommand{\arraystretch}{1.6}
        \begin{tabular}{llll}
        \hline
        \textbf{Setting} & \textbf{$Z$} & \textbf{$M$} & \textbf{$Y$}\\
        \hline
        1.1 & Continuous & Continuous & Continuous \\
        2.1 & Continuous & Continuous & Binary \\
        3.1 & Continuous & Binary & Continuous \\
        4.1 & Continuous & Binary & Binary \\
        1.2 & Binary & Continuous & Continuous \\
        2.2 & Binary & Continuous & Binary \\
        3.2 & Binary & Binary & Continuous \\
        4.2 & Binary & Binary & Binary \\
        \hline        
        \end{tabular}
    \end{table}
\end{center}
Data-generating mechanisms are summarized in Table \ref{tab:mecha}.
\begin{center}
    \begin{table}[H]
        \centering
        \caption{Data generating mechanism}
        \label{tab:mecha}
        \renewcommand{\arraystretch}{1.6}
        \begin{tabular}{p{2.5cm} p{2.5cm} p{10cm}}
        \hline
        \textbf{Variable} & \textbf{Type} & \textbf{Generating mechanism}\\
        \hline
        \multirow{2}{*}{Z} & Continuous & $\left(\begin{array}{c} Z(0) \\ Z(1) \end{array}\right) \sim \mathcal{N}\left[ \left( \begin{array}{c} 0.5 \\ 0.5+\alpha_1 \end{array} \right), \left(\begin{array}{cc}
            1 & \rho \\
             \rho & 1
        \end{array}\right)\right] $\\
        & Binary & $Z(a) \sim \text{Bern}(\text{expit}(0.5 + \alpha_1 a))$\\
        \hline
        \multirow{2}{*}{T} & Continuous &  $T(0) \sim \mathcal{N}(0.5,1)  \qquad \mathrm{and} \qquad T(1) \sim \mathcal{N}(0.5+\alpha_1,1)$ \\
        & Binary & $T(0) \sim \text{Bern}(0.5) \qquad \mathrm{and} \qquad T(1) \sim \text{Bern}(0.5+\alpha_1)$\\
        \hline
        
        \multirow{2}{*}{M} & Continuous & $M(a,z) = 0.5 + 0.5 a + \beta_2 z + 2.0 az + \varepsilon_M, \quad \varepsilon_M \sim \mathcal{N}(0, 1)$\\
        & Binary & $M(a,z) \sim \text{Bern}(\text{expit}(0.5 + 0.5a + \beta_2z + 2.0az))$\\
        \hline

        \multirow{2}{*}{Y} & Continuous & $Y(a,z,m) = 1.5 + 1.5 a + 1.5 z + 1.5 m  + 1.0 az + 1.0 am + \gamma_6 zm + 1.0 azm + \varepsilon_Y, \quad \varepsilon_Y \sim \mathcal{N}(0, 1)$\\
        & Binary & $Y(a,m,z) \sim \text{Bern}(\text{expit}(1.5 + 1.5a + 1.5z+ 1.5m + 1.0az+ 1.0am + \gamma_6zm + 1.0azm))$\\
        \hline
        
        \end{tabular}
    \end{table}
\end{center}
The key parameters $\rho$, $\alpha_1$, $\beta_2$ and $\gamma_6$ can take different values as follow:
\begin{align*}
    \rho &= \begin{pmatrix} -0.75& -0.20& 0.20& 0.75 \end{pmatrix}\\
    \alpha_1 &= \begin{pmatrix} -1.50& -0.50& 0.50& 1.50 \end{pmatrix}\\
    \beta_2 &=\begin{pmatrix} -1.50& -1.00& 1.00& 1.50 \end{pmatrix}\\
    \gamma_6 &= \begin{pmatrix} -1.50& -1.00& 0.00 & 1.00& 1.50 \end{pmatrix}
\end{align*}
Results of settings 1.2, 2.2, and 3.1 to 4.2 are presented in Figure \ref{fig:a6} to \ref{fig:a11}. Across all settings, when the natural and recanting-twin effects have opposite directions (i.e. $\psi_{P_2}^{ne} \psi_{P_2}^{rt} <0$ for path $P_2$ and $\psi_{P_3}^{ne} \psi_{P_3}^{rt} <0$ for path $P_3$), we evaluated the magnitude of the absolute difference between the two types of effect. Results of this assessment are summarized in Figure \ref{fig:a5}.

\newpage
\begin{landscape}    
    \begin{figure}
        \centering
        \includegraphics[width=0.7\linewidth]{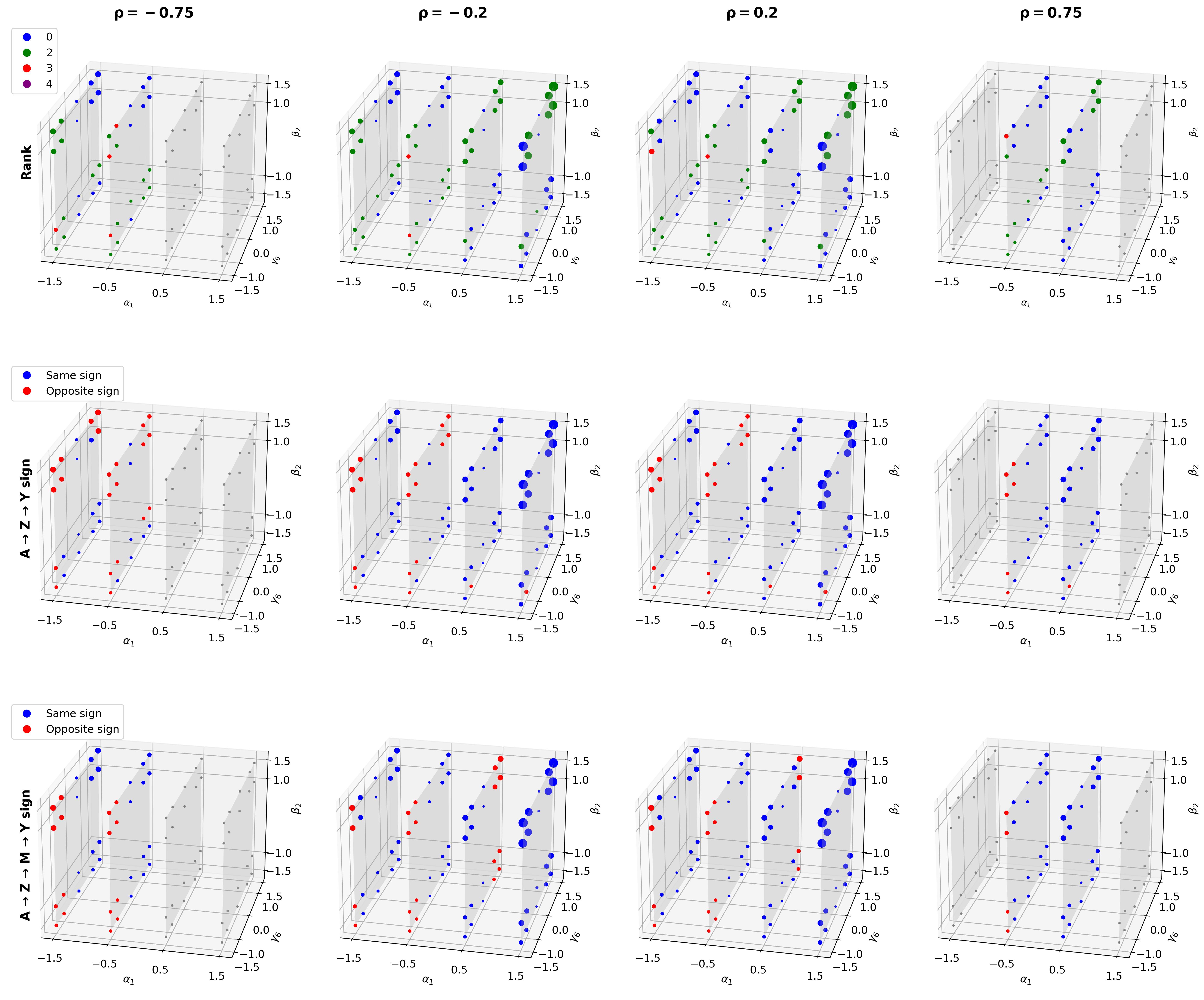}
        \caption{Disagreement in ranking and sign of natural and recanting-twin PSEs in setting 1.2 (binary $Z$, continuous $M$ and $Y$). First panel: number of paths among $P_j:j=1,\ldots, 4$ whose relative ranking differ under the natural versus recanting-twin system. Second panel: red and blue dots represent settings with disagreement and agreement in sign between $\psi_{P_2}^{ne}$ and $\psi_{P_2}^{rt}$, respectively. Third panel: red and blue dots represent settings with disagreement and agreement in sign between $\psi_{P_3}^{ne}$ and $\psi_{P_3}^{rt}$, respectively. The size of the dots is relative to the magnitude of the intermediate confounding parameter $\psi_{P_2\vee P_3}$.}
        \label{fig:a6}
    \end{figure}
    
    \begin{figure}
        \centering
        \includegraphics[width=0.7\linewidth]{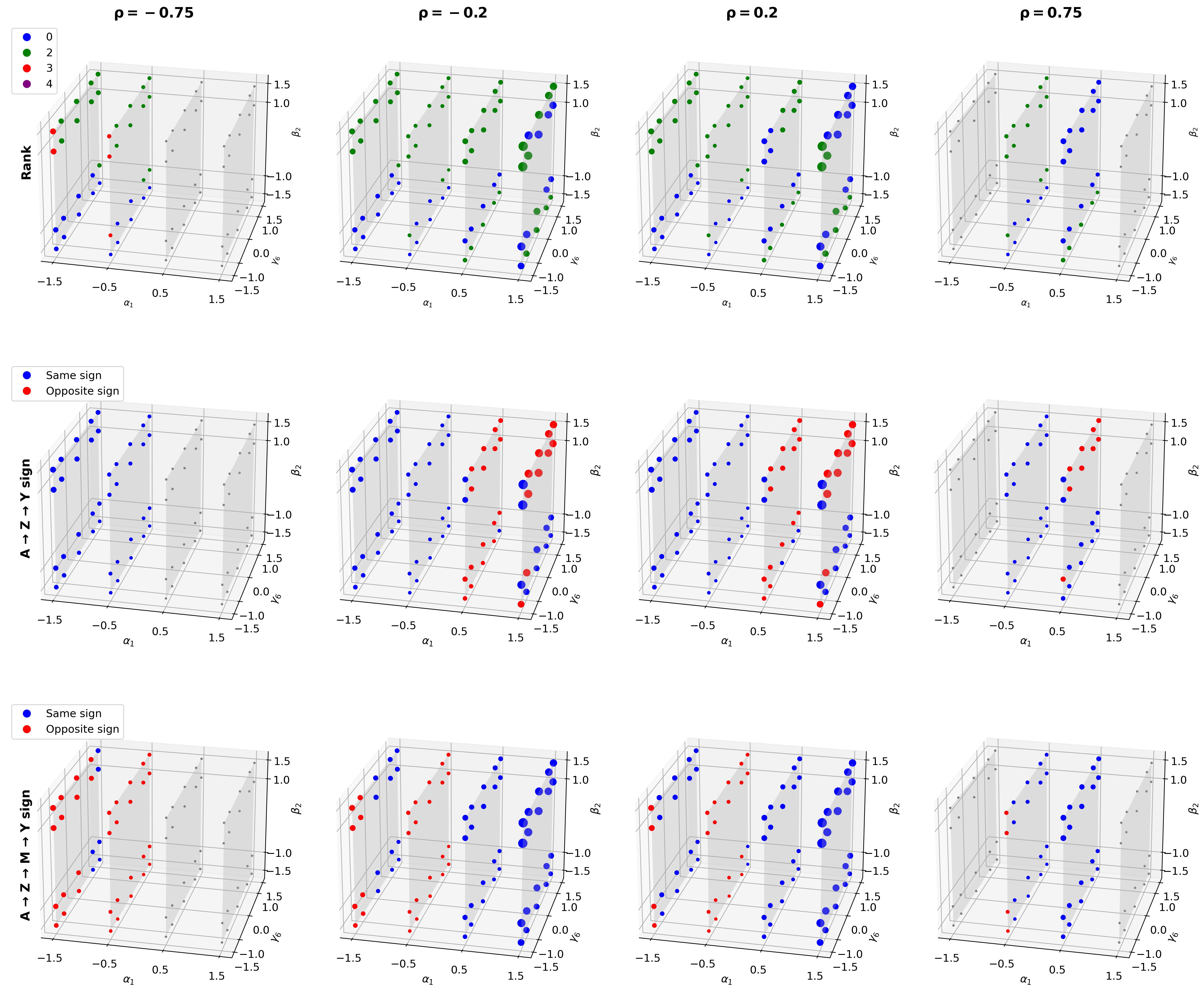}
        \caption{Disagreement in ranking and sign of natural and recanting-twin PSEs in setting 2.2 (binary $Z$ and $Y$, continuous $M$). First panel: number of paths among $P_j:j=1,\ldots, 4$ whose relative ranking differ under the natural versus recanting-twin system. Second panel: red and blue dots represent settings with disagreement and agreement in sign between $\psi_{P_2}^{ne}$ and $\psi_{P_2}^{rt}$, respectively. Third panel: red and blue dots represent settings with disagreement and agreement in sign between $\psi_{P_3}^{ne}$ and $\psi_{P_3}^{rt}$, respectively. The size of the dots is relative to the magnitude of the intermediate confounding parameter $\psi_{P_2\vee P_3}$.}
        \label{fig:a7}
    \end{figure}
    
    \begin{figure}
        \centering
        \includegraphics[width=0.7\linewidth]{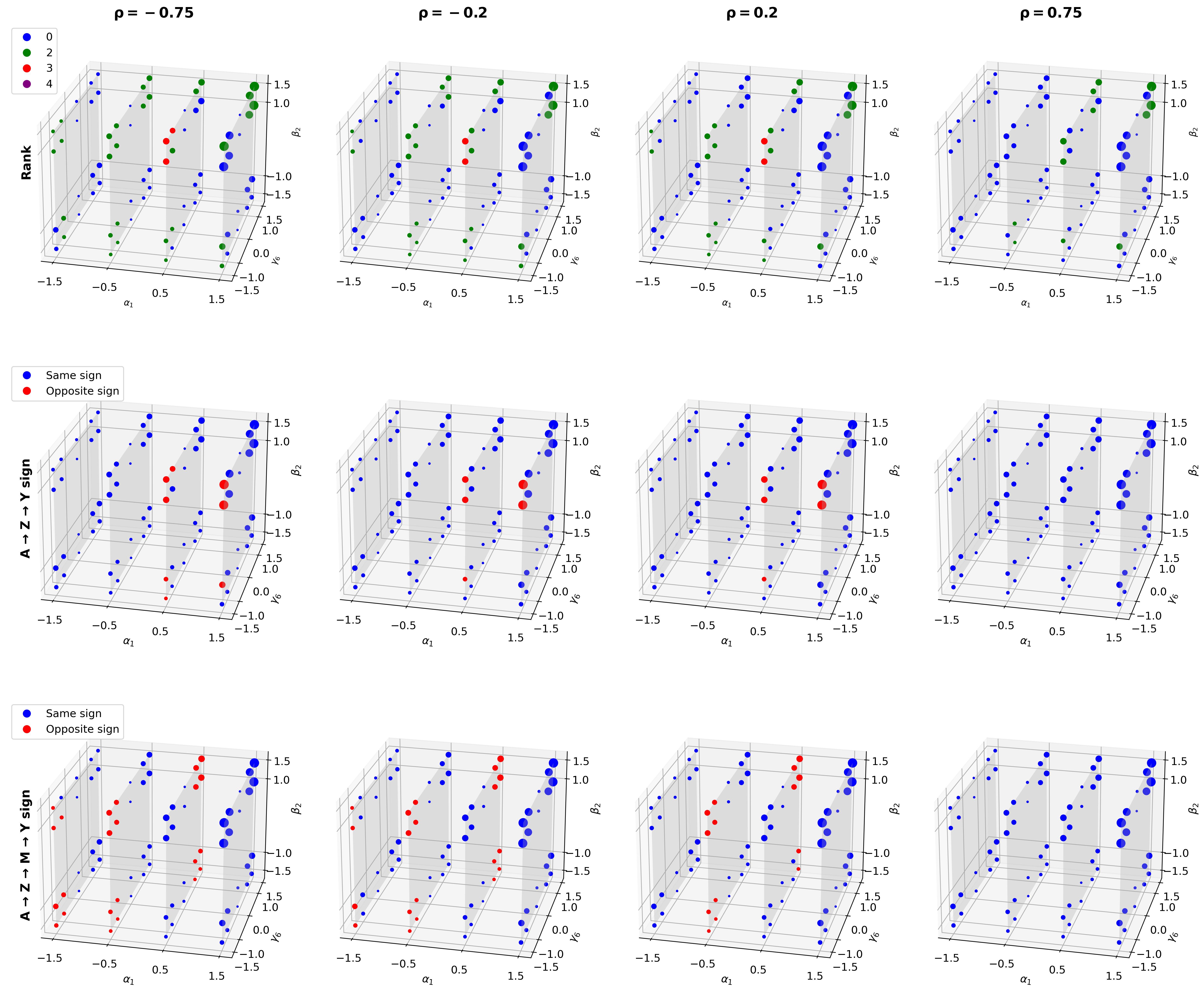}
        \caption{Disagreement in ranking and sign of natural and recanting-twin PSEs in setting 3.1 ($Y$ and $Z$ continuous, $M$ binary). First panel: number of paths among $P_j:j=1,\ldots, 4$ whose relative ranking differ under the natural versus recanting-twin system. Second panel: red and blue dots represent settings with disagreement and agreement in sign between $\psi_{P_2}^{ne}$ and $\psi_{P_2}^{rt}$, respectively. Third panel: red and blue dots represent settings with disagreement and agreement in sign between $\psi_{P_3}^{ne}$ and $\psi_{P_3}^{rt}$, respectively. The size of the dots is relative to the magnitude of the intermediate confounding parameter $\psi_{P_2\vee P_3}$.}
        \label{fig:a8}
    \end{figure}

    \begin{figure}
        \centering
        \includegraphics[width=0.7\linewidth]{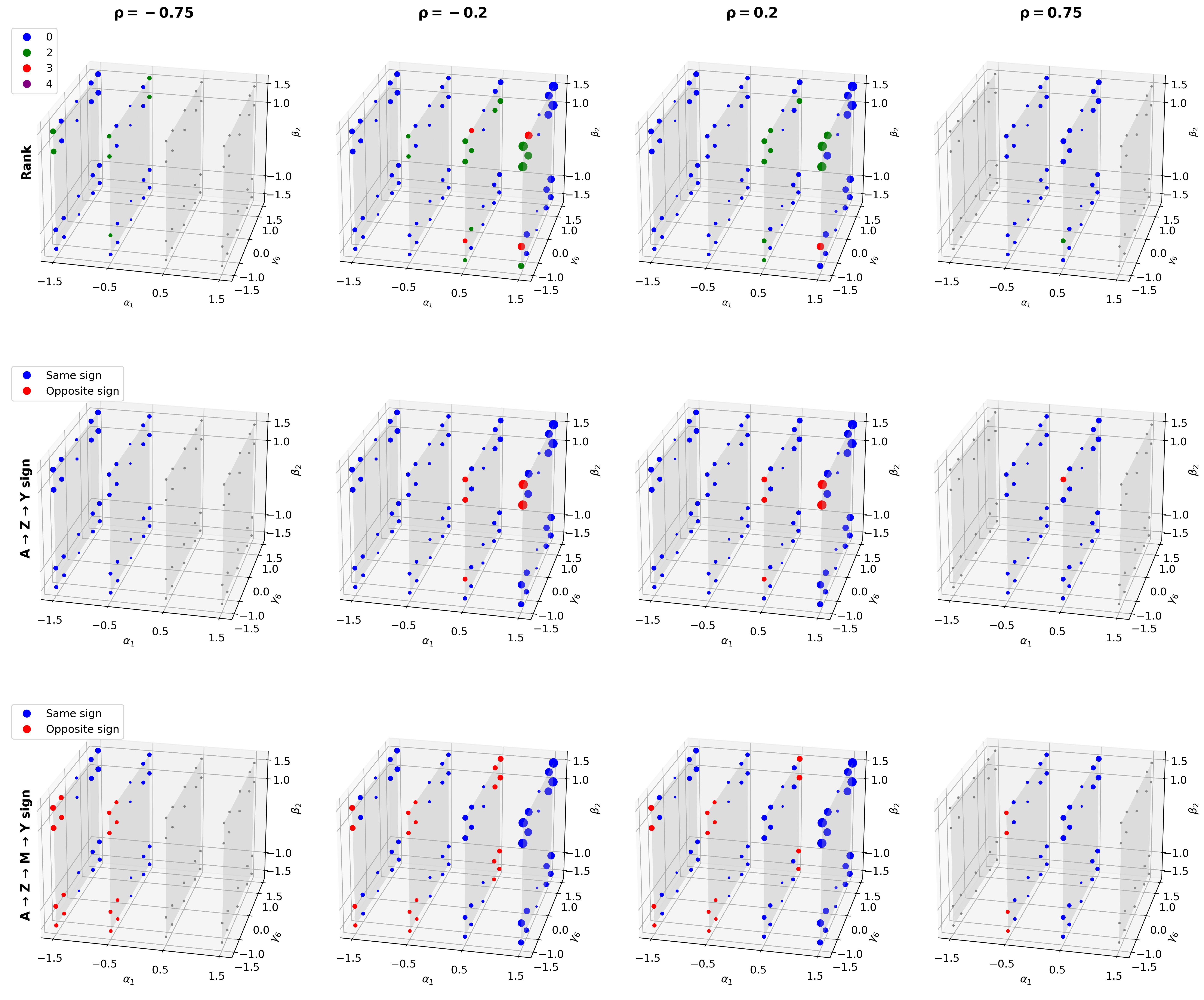}
        \caption{Disagreement in ranking and sign of natural and recanting-twin PSEs in setting 3.2 ($Y$ continuous, $Z$ and $M$ binary). First panel: number of paths among $P_j:j=1,\ldots, 4$ whose relative ranking differ under the natural versus recanting-twin system. Second panel: red and blue dots represent settings with disagreement and agreement in sign between $\psi_{P_2}^{ne}$ and $\psi_{P_2}^{rt}$, respectively. Third panel: red and blue dots represent settings with disagreement and agreement in sign between $\psi_{P_3}^{ne}$ and $\psi_{P_3}^{rt}$, respectively. The size of the dots is relative to the magnitude of the intermediate confounding parameter $\psi_{P_2\vee P_3}$.}
        \label{fig:a9}
    \end{figure}

    \begin{figure}
        \centering
        \includegraphics[width=0.7\linewidth]{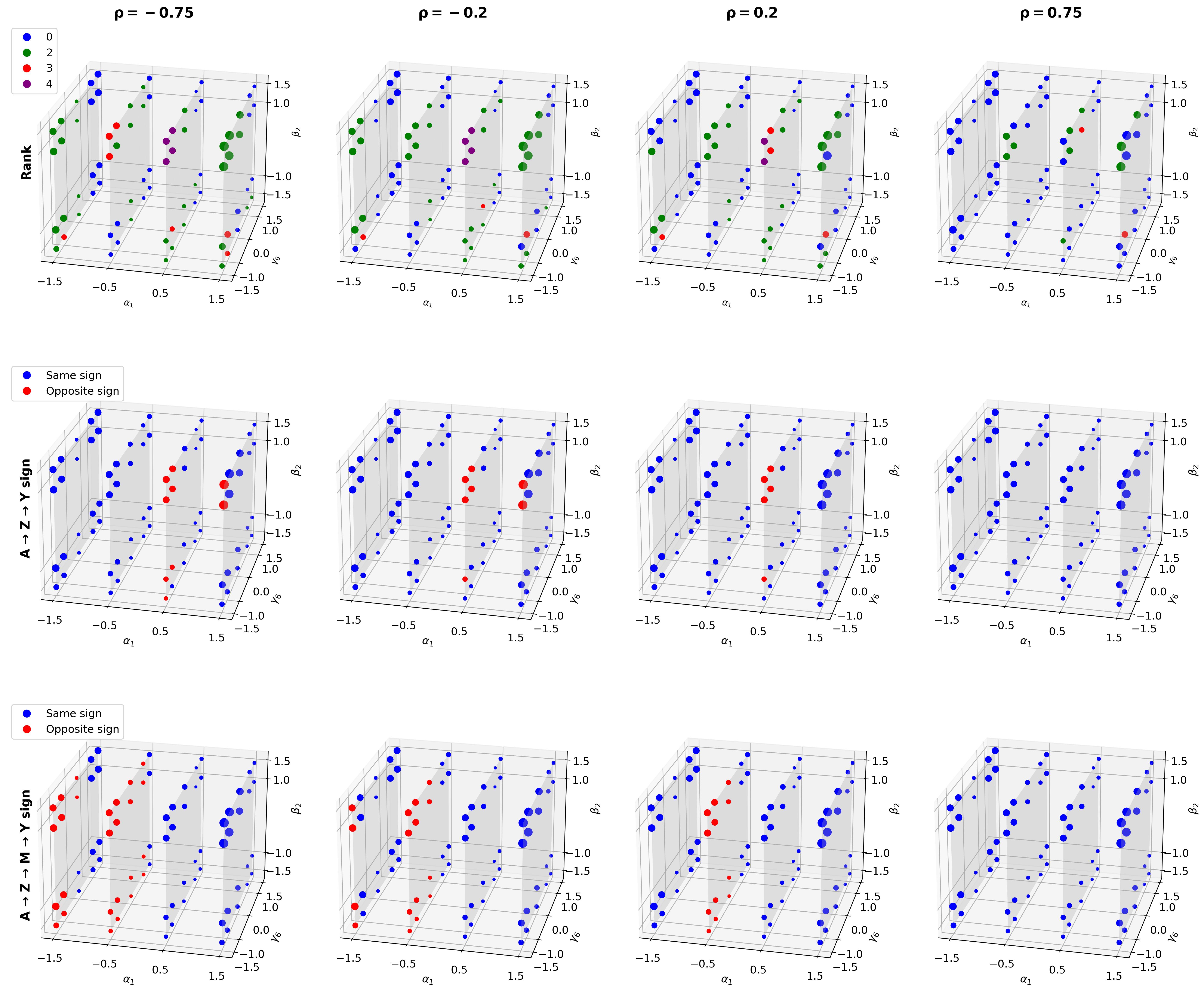}
        \caption{Disagreement in ranking and sign of natural and recanting-twin PSEs in setting 4.1 ($Y$ binary, $Z$ continuous, $M$ binary). First panel: number of paths among $P_j:j=1,\ldots, 4$ whose relative ranking differ under the natural versus recanting-twin system. Second panel: red and blue dots represent settings with disagreement and agreement in sign between $\psi_{P_2}^{ne}$ and $\psi_{P_2}^{rt}$, respectively. Third panel: red and blue dots represent settings with disagreement and agreement in sign between $\psi_{P_3}^{ne}$ and $\psi_{P_3}^{rt}$, respectively. The size of the dots is relative to the magnitude of the intermediate confounding parameter $\psi_{P_2\vee P_3}$.}
        \label{fig:a10}
    \end{figure}

    \begin{figure}
        \centering
        \includegraphics[width=0.7\linewidth]{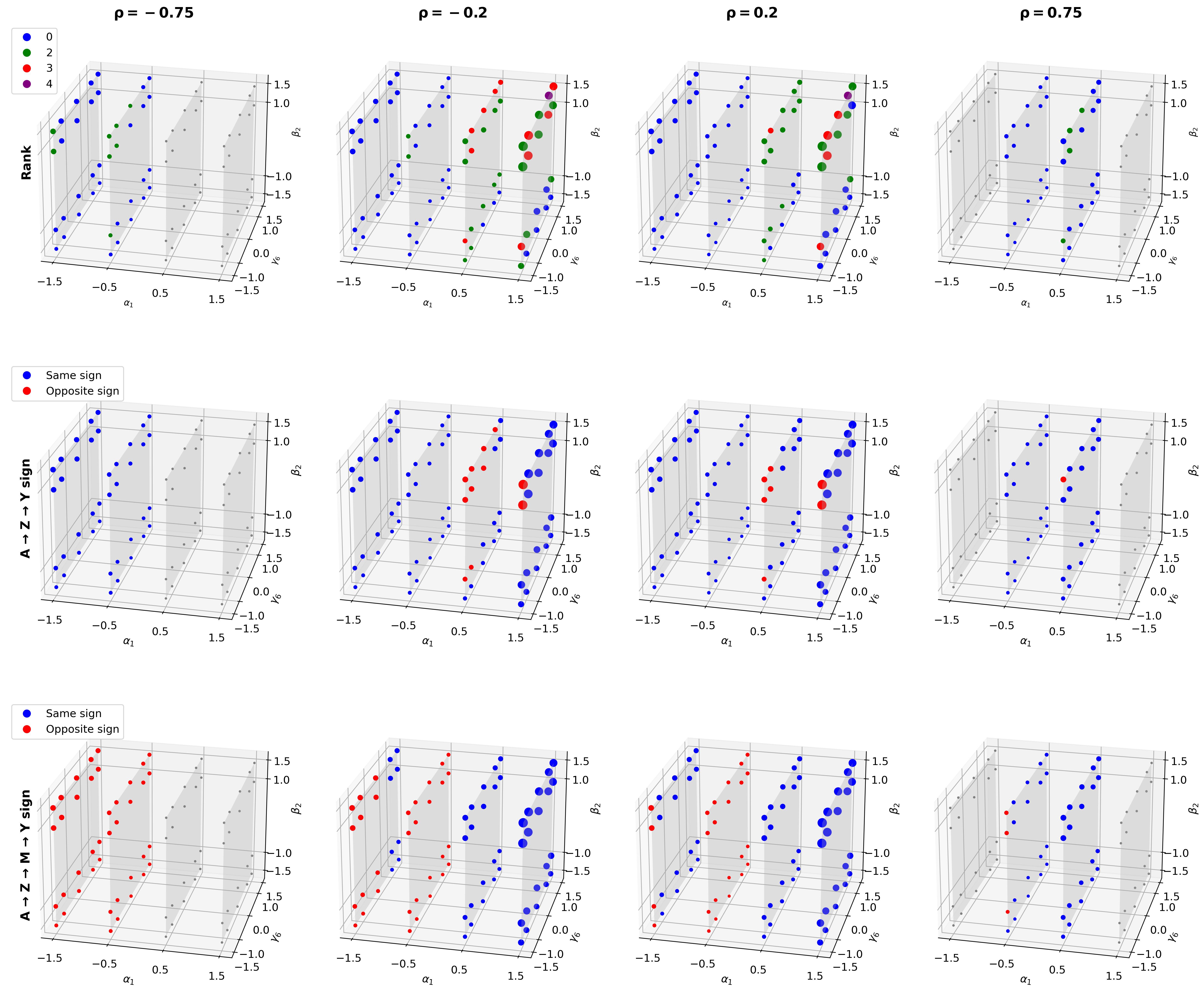}
        \caption{Disagreement in ranking and sign of natural and recanting-twin PSEs in setting 4.2 ($Y, Z, M$ all binary). First panel: number of paths among $P_j:j=1,\ldots, 4$ whose relative ranking differ under the natural versus recanting-twin system. Second panel: red and blue dots represent settings with disagreement and agreement in sign between $\psi_{P_2}^{ne}$ and $\psi_{P_2}^{rt}$, respectively. Third panel: red and blue dots represent settings with disagreement and agreement in sign between $\psi_{P_3}^{ne}$ and $\psi_{P_3}^{rt}$, respectively. The size of the dots is relative to the magnitude of the intermediate confounding parameter $\psi_{P_2\vee P_3}$.}
        \label{fig:a11}
    \end{figure}
\end{landscape}
\end{document}